\documentclass[journal]{vgtc}                      
\usepackage{xfrac}
\usepackage{multirow}
\usepackage{amssymb}
\usepackage{enumitem}
\usepackage{dblfloatfix}
\onlineid{1622}
\vgtccategory{Research}
\vgtcpapertype{algorithm/technique}
\newcommand{\paptit}{CrossSet: Unveiling the Complex Interplay of Two Set-typed Dimensions in Multivariate Data}
\title{\paptit}
\author{%
\authororcid{Kre{\v{s}}imir Matkovi{\'{c}}}{0000-0001-9406-8943},~ 
{Rainer Splechtna},~
\authororcid{Denis Gra{\v{c}}anin}{0000-0001-6831-2818},~ 
and \authororcid{Helwig Hauser}{0000-0003-0395-3192}
}
\authorfooter{
\item
Kre{\v{s}}imir Matkovi{\'{c}} and Rainer Splechtna are with the VRVis Research Center in Vienna, Austria; eMail: \{\,Matkovic\,|\,Splechtna\,\}@VRVis.at.
\item
Denis Gra\v{c}anin is with Virginia Tech; eMail: gracanin@vr.edu.
\item
Helwig Hauser is with the University of Bergen in Norway; eMail: Helwig.Hauser@UiB.no.
}
\abstract{%
The interactive visual analysis of set-typed data, i.e., data with attributes that are of type \textit{set}, is a rewarding area of research and applications.
Valuable prior work has contributed solutions that enable the study of such data with individual set-typed dimensions.
In this paper, we present CrossSet, a novel method for the joint study of two set-typed dimensions and their interplay.  
Based on a task analysis, we describe a new, multi-scale approach to the interactive visual exploration and analysis of such data.  
Two set-typed data dimensions are jointly visualized using a hierarchical matrix layout, enabling the analysis of the interactions between two set-typed attributes at several levels, in addition to the analysis of individual such dimensions.  
CrossSet is anchored at a compact, large-scale overview that is complemented by drill-down opportunities to study the relations between and within the set-typed dimensions, enabling an interactive visual multi-scale exploration and analysis of bivariate set-typed data. 
Such an interactive approach makes it possible to study single set-typed dimensions in detail, to gain an overview of the interaction and association between two such dimensions, to refine one of the dimensions to gain additional details at several levels, and to drill down to the specific interactions of individual set-elements from the set-typed dimensions. 
To demonstrate the effectiveness and efficiency of CrossSet, we have evaluated the new method in the context of several application scenarios. 
}
\graphicspath{{figs/}{figures_revision_revision/}{pictures/}{images/}{./}} 
\usepackage{tabu}                      
\usepackage{booktabs}                  
\usepackage{lipsum}                    
\usepackage{mwe}                       
\usepackage{mathptmx}                  
\begin{document}
%
%
%
\firstsection{Introduction}\label{sec:intro}\maketitle
When dealing with set-typed data, we differentiate two complementary perspectives in terms of how sets are considered.
On the one hand, we can see subsets of data items, identified by shared data properties, such as all \textit{mammals} within a dataset of \textit{animals}. 
On the other hand, we can have set-typed data attributes, where the value of such an attribute is a set (per item)~-- the attribute \textit{academic degrees} in a dataset of employees is an example. 
At large, both perspectives can be mapped to each other~-- consider the subset of \textit{employees with a Master's degree}. 
In terms of data analysis, however, it makes a lot of sense to consider them as complementary: 
``\textit{Which items} have a particular property?'' 
vs.
``\textit{Which properties} does a particular item have?''. 
In the following, we consider data in the form of a data table~-- rows (data items) and columns (data attributes)~-- while focusing on cases where (at least) two of the attributes are of type \textit{set}.  
For instance, we could have patients with \textit{symptoms} and \textit{diseases}, i.e., two set-typed patient attributes.%
\begin{figure}[b!] 
\centering
\includegraphics[width=.925\columnwidth]{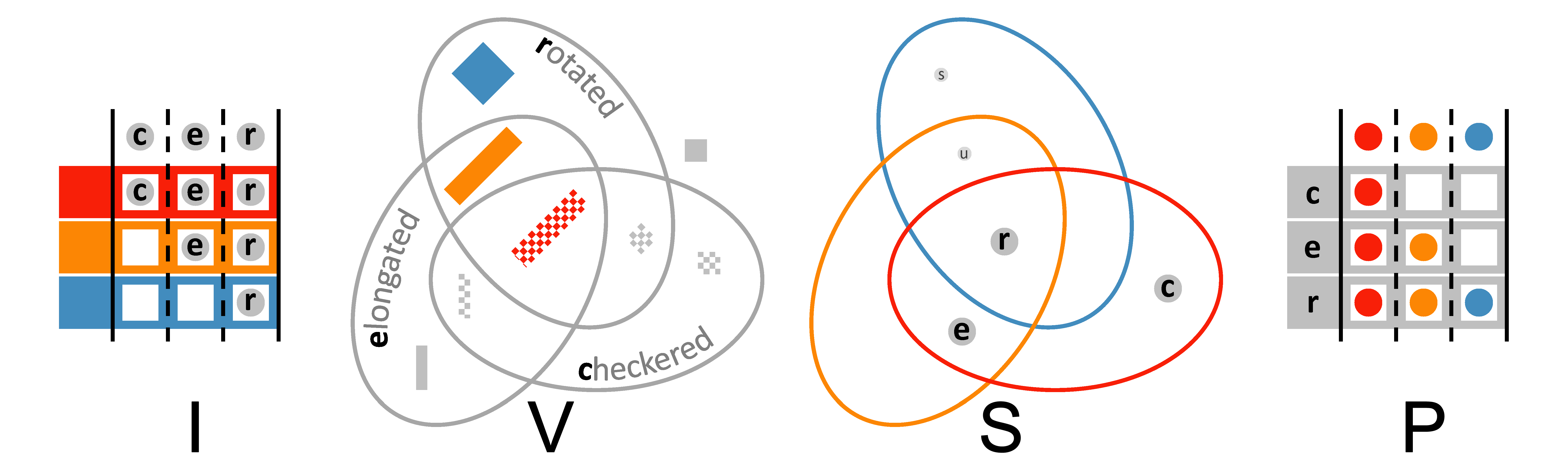}
\caption{Two complementary perspectives on set-typed data, illustrated by items ``red'', ``orange'', and ``blue'', each having a certain subset of the  properties \textbf{c}heckered, \textbf{e}longated, and \textbf{r}otated;  
here, item ``red''  has all three properties, while  ``blue'' is only \textbf{r}otated (\textbf{I}). 
The Venn-diagram~\textbf{V} illustrates the $2^3$ possible cases of which subset of properties an item can have, with ``orange'' placed in the intersection of \textbf{e}longated and \textbf{r}otated, but outside of \textbf{c}heckered; further possible items, not included in table~\textbf{I}, are added in gray. 
Alternatively to asking ``Which properties per item?'', we can ask ``Which items per property?''.
Properties-table \textbf{P} lists the items per property. 
In \textbf{S}, \textbf{e}longated is placed on ``red'' \& ``orange'', but not on ``blue''. 
Both perspectives are dual\,/\,complimentary to each other; \mbox{\textbf{V} scales} better for many items, while \textbf{S} scales better for many properties.%
} 
\label{fig:twopersp}
\end{figure}
\par 
A prominent share of related prior visualization research is rooted in the principal of Venn~\cite{Venn_1880} and Euler~\cite{euler1802letters} diagrams. 
Visual marks, representing items, are placed on a ``map'' of item subsets and their intersections (illustrated in Fig.~\ref{fig:twopersp}:V). 
Listing all items with a certain property (Fig.~\ref{fig:twopersp}:P) provides an immediate answer to the question ``\textit{Which items} have a particular property?''. 
In the item-centric perspective, however, we record~-- in a set-typed attribute~-- all properties per item (Fig.~\ref{fig:twopersp}:I). 
This enables to link this perspective with other item-based visualization approaches as we can naturally complement one such set-typed attribute with others, including more typical (quantitative, qualitative) attributes as well as additional set-typed attributes.
\par
While we have seen~\cite{Matkovic-2020-c,SetOGram2008,Alsallakh-2013-a} that properly addressing one set-typed attribute is already a substantial challenge, we clearly lack appropriate solutions for the relevant case of data with more than just one such set-typed attribute. 
%
While set-typed data attributes are not as common as regular quantitative and qualitative data attributes, we still need the same ``battery'' of exploration and analysis methods when facing such data, in particular also in the common multivariate case. 
\par
\begin{figure*}[t!] 
\centering
\includegraphics[width=\linewidth]{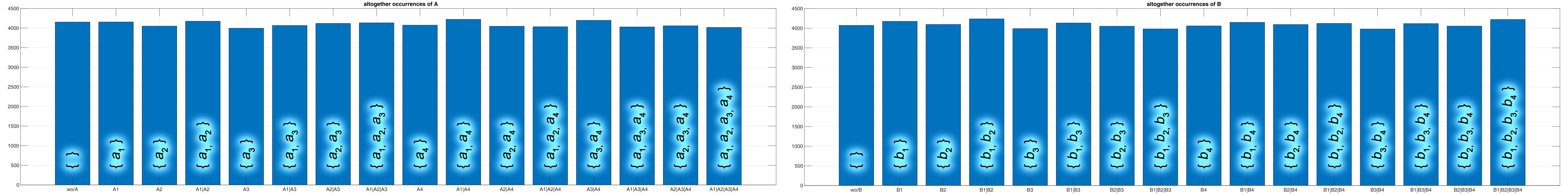}\\[-1ex]
\caption{Histograms showing the occurrences of all $2^4=16$ possible subsets of $\left\{a_1,\,a_2,\,a_3,\,a_4\right\}$ as \textbf{A}-values per data item, and of $\left\{b_1,\,b_2,\,b_3,\,b_4\right\}$ for \textbf{B}, in dataset \textbf{S1}~-- considering the two set-typed data dimensions \textbf{A} and \textbf{B} individually does not reveal the co-occurrence structure between \textbf{A} and \textbf{B}.} 
\label{fig:histAB}
\end{figure*}

Prior work on studying individual set-typed data attributes provides good solutions for the univariate exploration and analysis of such data.
With this paper, we contribute a solution for the bivariate exploration and analysis of multivariate data with two set-typed data attributes, providing valuable insight into the interplay of two such data attributes, especially when the bivariate association structure is impossible to see when only considering the set-typed data attributes individually, i.e., with univariate analysis.  To illustrate such a situation, we have constructed six synthetic datasets (\textbf{S1}\,\dots\,\textbf{S6}), which all look indistinguishable when only considering their two set-typed attributes \textbf{A} and \textbf{B} individually.  The data items in these datasets all have a particular subset of $\left\{a_1,\,a_2,\,a_3,\,a_4\right\}$ as entry in column \textbf{A} and a particular subset of $\left\{b_1,\,b_2,\,b_3,\,b_4\right\}$ as their set-typed \textbf{B}-value. In all of the six datasets, all possible subsets occur equally likely across all data items (Fig.~\ref{fig:histAB}).  
\begin{figure*}[t!] 
\centering
\includegraphics[width=\linewidth]{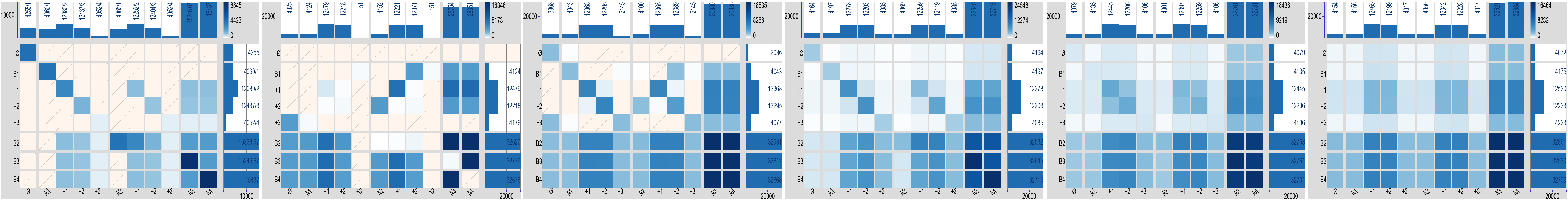}\\[-1ex]
\caption{Illustration that CrossSet shows visual differences, i.e., that it can unveil
different co-occurrence structures, for the six synthetic datasets \textbf{S1}, \textbf{S2}, \ldots, \textbf{S6} (left-to-right): ``one-to-one'' (\textbf{S1}), ``one-to-not-one'' (\textbf{S2}), 50\% ``one-to-one'' \& 50\% ``one-to-not-one'' (\textbf{S3}), 50\% ``one-to-one'' (\textbf{S4}), 12.5\% ``one-to-one'' (\textbf{S5}), all random (\textbf{S6}). More details about the data in the text and an expanded version of this figure in Fig.~\ref{fig:probdemoCrossSet02} (suppl.~material).} 
\label{fig:synthCS}
\end{figure*}
\par
Even though all six datasets show the same, even 1D distribution for both \textbf{A} and \textbf{B}~-- according histograms look all similar as shown in Fig.~\ref{fig:histAB}~-- we still realized significantly different co-occurrence structures:~ 
\mbox{in \textbf{S1}}, \textbf{A}-subsets co-occur with \textbf{B}-subsets ``one-to-one''~-- example: if $\left\{a_1,\,a_4\right\}$ is the \textbf{A}-value for data item $i$, then its \textbf{B}-value is $\left\{b_1,\,b_4\right\}$~--,~ 
\mbox{in \textbf{S2}}, \mbox{\textbf{A}-subsets} co-occur with \textbf{B}-subsets ``one-to-not-one'' (at large), i.e., $\left\{a_1\right\}$ would co-occur with $\left\{b_2,\,b_3,\,b_4\right\}$,~ 
\mbox{in \textbf{S3}}, for half of the items, \textbf{A}-values co-occur with \textbf{B}-values ``one-to-one'', while for the other $50\%$, they co-occur ``one-to-not-one'',~ 
\mbox{in \textbf{S4}}, for half of the data items there's a ``one-to-one'' co-occurrence, while for the other $50\%$, there is none (random co-occurrence),~ 
\mbox{in \textbf{S5}}, this ``one-to-one'' co-occurrence is given for $12.5\%$ of the data,~ 
and \textbf{S6} only has random co-occurrence.  
%
None of the so far published visualization methods for set-typed data can reveal the different co-occurrence structures in \textbf{S1}, \textbf{S2}, \ldots, \textbf{S6}~as compactly illustrated with the help of CrossSet in Fig.~\ref{fig:synthCS}~-- at least not without substantial additional interaction and linking and brushing.  
\par  
To come up with a useful solution, we first considered the following:%
\par
(1)~All the valuable previous work on analyzing individual set-typed data attributes amounts to an important basis of CrossSet, including both the questions asked, when facing one set-typed data attribute, as well as the according design choices (see Section~\ref{sec:related_work} on related work).%
\par
(2)~Further, we rest our study on typical data analysis trajectories: data scientists usually investigate simple data properties first, before engaging with more intricate aspects; in particular, it is common to first study the data dimensions individually, before analyzing their relation.%
\par 
(3)~We also appreciate that set-typed data attributes share important commonalities with qualitative data~-- in particular with Boolean data attributes as we can think of attribute properties as either being present, or not; often such data are dealt with by counting (say, in a histogram, or in a contingency table, etc.).
\par
One obvious default consideration, when attempting the bivariate analysis of set-typed data, is to map the problem to a domain for which valuable approaches exist.
In our case, one such mapping would be to treat all possible subsets as unique, categorical values.
In statistics, this is often done by use of contingency tables which show counts of different associations~\cite{agresti:categorical}. 
In visualization, this would open up for available solutions that cope with categorical data, including histograms for the univariate case and Mosaic plots~\cite{Mosaic1981} as well as parallel sets~\cite{par-sets} for the multivariate case.  
\par
Unfortunately, this leads to (at least) two critical problems:
\par 
(1)~This way, we would lose the important set-semantic of the data~-- without a major work-around we would be unable to straight-forwardly relate to the size of the subset, for example. 
This approach would also require a particular ordering of all subsets and any one of such orderings would be arbitrary among exponentially many possible choices.  
\par 
(2)~Besides this principal problem, we also would quickly run into enormous technical challenges due to the vast size of these power sets~-- most visualization techniques, that consider categories as first-order entity, cannot deal with hundreds of them (or many, many more).  
\par
Therefore, we concluded that a new approach is needed, in particular when aiming at a bivariate analysis of data with set-typed dimensions.
We found it instructive to extrapolate from a similar reasoning about the extension from univariate to bivariate data analysis (and beyond) for regularly-typed data, where the main point is to get a grip on everything multi-dimensional that is not already provided by the Cartesian product of the univariate analysis of the individual dimensions.
%
Overall, we aimed at the following three main objectives: 
\par 
\textbf{Enable the study of truly two-dimensional data relations.} 
Non-separable 2D data structures cannot be identified and characterized without multivariate analysis methods~-- above we illustrated this with the six synthetic datasets \textbf{S1}, \textbf{S2}, \ldots, \textbf{S6}.
Therefore, our central objective was to design CrossSet to enable the identification and characterization of truly 2D relations between two set-typed data dimensions.  
\par 
\textbf{Tackle the inherently given scalability challenges.} 
We face the challenge of exponential growth in the number of possible subsets. 
Realistic scenarios can easily come with set-typed data attributes with large numbers of set elements per attribute. 
If only considered at the ``all details'' level, we quickly have hundreds (or even many more) counts to consider, causing serious scalability problems both in technical terms, as well as in perceptual and cognitive terms. 
We thus needed an appropriate multi-scale solution with suitable interaction.
\par 
\textbf{Support of an items-centric and a properties-centric analysis.} 
This objective relates to the two complementary perspectives on sets as described above. 
At times, it is important to focus on the set elements for answering relevant questions (in a hospital, for example, one may need to know how many occurrences of a disease there are to acquire the right number of drugs).
At other times, the main focus is on the data items as first-order objects of the analysis (other questions in the hospital relate to the number of patients, irrespective of how many diseases\,/\,symptoms they have, e.g., when accounting for a sufficient number of beds). 
Since much of the analysis is based on counting at different levels of aggregation, we needed to enable counting set elements as well as counting data items. 
\par
Our solution, CrossSet, combines a multi-scale visualization design  with appropriate interaction.  
In addition to new capabilities, oriented towards unveiling the complex interplay of two set-typed dimensions, CrossSet also provides means for the univariate analysis of individual set-typed attributes.
It is naturally linked with other views, integrating CrossSet with the interactive visual study of other data attributes, as well as with other, univariate approaches to the study of set-typed data (as, e.g., Set'o'grams~\cite{SetOGram2008}). 
\section{Related work}
\label{sec:related_work}
The problem of finding a suitable visualization for an effective and efficient exploration of set-typed data has been addressed repeatedly.  Alsallakh et al.~\cite{eurovisstar_2014,Alsallakh2016cgf} presented a systematic overview of state-of-the-art techniques for set visualization, describing commonly supported tasks and classifying the most often used visual representations into six categories. 
These include Euler and Venn diagrams~\cite{euler1802letters, Venn_1880,Baron-1969-a,UntanglingEuler_2010, Rodgers-2014-a},
overlays~\cite{Itoh-2009-a,Gansner-2010-a,Dinkla-2012-a,Meulemans-2013-a},
node-link diagrams~\cite{Kazuo-2006-a,Alsallakh-2013-a, UpSet_2014, OnSet_2014},
matrix-based techniques~\cite{Kim-2007-a, UpSet_2014, Yalcin-2016-a, GridSet_2020},
aggregation-based techniques~\cite{SetOGram2008, Alsallakh-2013-a,Gove-2014-a, UpSet_2014, OnSet_2014, Yalcin-2016-a, PowerSet_2017, RainBio_2018, GridSet_2020},
and other techniques~\cite{Dennig-2024-a,Li-2025-a,Mosaic1981, SetOGram2008, PowerSet_2017, LuzMasoodian_2018,time-sets_2018,Nur&Nur_2019,Piccolotto-2025-a,Stapleton_2019, GridSet_2020,Zhu-2023-a}. 
Due to the complexity of the problem and the broad variety of tasks, it is common to combine several techniques~\cite{SetOGram2008,Alsallakh-2013-a,UpSet_2014,OnSet_2014,Yalcin-2016-a,PowerSet_2017,GridSet_2020,Sohns-2022-a}.
\par 
Visualization design is guided by specific priorities in the application. 
A common choice is to use a matrix-based representation, due to its readability and intuitiveness.  
UpSet~\cite{UpSet_2014}, for example, uses a dotted matrix layout to represent the association between data items (rows) and sets (columns), and 
AggreSet~\cite{Yalcin-2016-a} uses a matrix plot to visualize pair-wise set intersections.  In CrossSet, heatmaps are used to depict the co-occurrence structure of two set-typed dimensions. 
Other techniques, such as RadialSet~\cite{Alsallakh-2013-a} and DualRadialSet~\cite{Matkovic-2020-c}, use a radial layout.  
In RadialSet~\cite{Alsallakh-2013-a}, set memberships are analysed using visual links.  
GridSet~\cite{GridSet_2020} also 
employs visual links to depict relations between elements across different sets, but does that in a matrix-like layout.
Tang et al.~\cite{Tang-2024-a} provide the set space for the multi-label attribute data and use an integrated, ordered radial coordinate system for set projection and color mapping.
RangeSets~\cite{Sohns-2022-a} use a set-based visualization approach for binned attribute values to observe structure and detect outliers.
\par 
Many approaches deploy frequency-based representations. 
Statistical aggregation at the element and attribute levels contributes to an insightful visualization and helps with scalability.
UpSet~\cite{UpSet_2014} supports the exploration of set interactions and set memberships of individual elements by task-driven aggregates based on groupings and queries. 
GridSet~\cite{GridSet_2020} and PowerSet~\cite{PowerSet_2017} use treemaps~\cite{shneiderman_2009} to provide a comprehensive overview of non-empty set intersections. 
AggreSet~\cite{Yalcin-2016-a} employs aggregations for individual data dimensions in forms of sets, set-degrees, set-pair intersections, and other attributes, with a focus on good scalability.
OnSet~\cite{OnSet_2014} enables the interactive visualization of large binary set data, which is based on modelling each set by the elements that it simultaneously contains in a single, combined domain of elements for all sets. 
For multiple set operations, GridSet~\cite{GridSet_2020} supports set operations by dragging and dropping of set objects.
\par
Most visualization techniques provide further details on demand.
UpSet~\cite{UpSet_2014} provides information at the element level in a separate view. 
A popular visualization design choice for statistical details on demand are histograms. 
AggreSet~\cite{Yalcin-2016-a} shows histograms of set-related and other attributes. 
In CrossSet, we also provide further statistical aggregations  (at different levels of detail) along with interactive information drill down and details on demand.
\par 
Another recent approach uses biclustering as a fundamental data mining technique for a coordinated relationship analysis in large textual datasets. 
Examples include BiDots~\cite{Zhao-2018-a}, Bixplorer~\cite{Fiaux-2013-a}, Five-Level Design Framework for Bicluster Visualizations~\cite{Sun-2014-a}, and Bi-Set~\cite{Sun-2016-a}. 
Data subsets with similar properties are extracted and groups of closely related entities detected. 
Although these visualizations were initially designed for analyzing relational subsets of the data, they do address similar tasks and follow similar approaches when aggregating and visualizing set relationships. 
In general, these techniques show the relationship between single elements and the elements of a single set-typed attribute, 
as well as their distributions~\cite{Theus-2008-a, Mondrian, Weaver-2010-a}. 
\par
In the case of two set-typed data dimensions, it is possible to deploy two instances of an univariate method in a linked system, providing only limited opportunities for a multivariate analysis. 
Using two Set'o'grams~\cite{SetOGram2008}, for example, in a coordinated multiple views system, can reveal separable relations in the data.
DualRadialSet~\cite{Matkovic-2020-c} is an extension of RadialSet~\cite{Alsallakh-2013-a} and the first method that simultaneously  visualizes two set-typed dimensions to analyze relations between data elements from different set-typed attributes. 
It is, however, limited to a high-level aggregation and does not allow a detailed bivariate analysis.  
CrossSet shows the relationship between two set-typed data dimensions, both as high-level overview and as low-level details. 
To our knowledge, CrossSet is the first visualization technique that provides thorough means for a bivariate exploration and analysis of data with (at least) two set-typed dimensions.
\section{Tasks and design requirements} 
\label{sec:analysis}
In this section, we reason about supported tasks and according design requirements for an effective approach to the exploration and analysis of data with two set-typed attributes. 
During data and task abstraction, we understood that a layered aggregation strategy was needed to cope with the inherently given scalability challenges. In particular, we identified the need for the following two approaches:
\par 
(a)~We found that it was common in multiple analysis scenarios to ``pick'' individual set elements as ``anchor point'' from which one  then would relate to others.
When studying patient data, for example, it is natural to start with one symptom\,/\,disease and then see, how other symptoms\,/\,diseases relate.  
\par 
(b)~We also found that frequently the count of set elements~-- per set-typed attribute value~-- was a starting point for drilling into more details. 
For instance, one would analyze how patients with several diseases compare to others with only a few.  
\begin{figure*}[b!]
\vspace{-1ex}\centering 
\includegraphics[width=\linewidth]{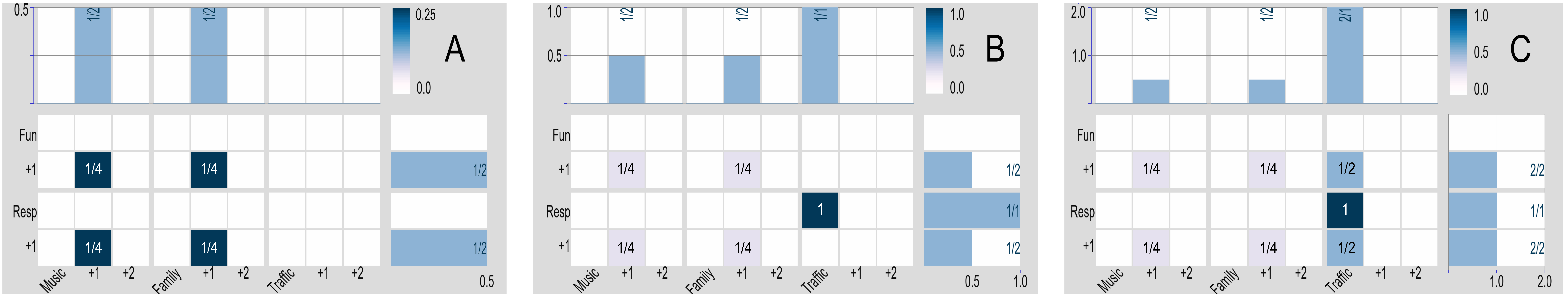}\\[-1ex]
\caption{Basic visual encoding of CrossSet:~ 
(A)~A data line with $\{\textit{Music},\,\textit{Family}\}$ as \textit{Input}-value and $\{\textit{Fun},\,\textit{Resp}\}$ as \textit{Output}-value;~ 
(B)~A 2\textsuperscript{nd} data line added with $\{\textit{Traffic}\}$ as \textit{Input}-value and $\{\textit{Resp}\}$ as \textit{Output}-value;~ 
(C)~A 3\textsuperscript{rd} data line with $\{\textit{Traffic}\}$ as \textit{Input}-value and $\{\textit{Fun},\,\textit{Resp}\}$ as \textit{Output}-value.}
\label{fig:basicPrinciple1}
\end{figure*}
\par\smallskip 
At large, we identified the following high-level tasks in the bivariate exploration and analysis of set-typed data: 
\vspace{0ex}\begin{itemize}\setlength\itemsep{-0.5ex} 
\item \textbf{T1: analyzing co-occurrence patterns.} 
Clearly, we aim at identifying and characterizing relevant patterns of co-occurrence, studying how entries of one set-typed dimension co-occur with the entries of another one. Different types of co-occurrence are relevant, as well as the related set cardinalities. A one-to-one co-occurrence structure, for example, is a special case where a certain set-typed value of one dimension is one-to-one related to a certain set-typed value of the second dimension (dataset \textbf{S1}, for example). The cardinalities of the involved set-typed values make a difference, as well. Similarly, one-to-many and many-to-many co-occurrences are also important to be identified and analyzed. 
\item \textbf{T2: studying outliers.} 
Complementary to the analysis of patterns~(T1), the study of outliers is also common and important: Are there special\,/\,atypical co-occurrences in the data?  In parallel to revealing patterns at different levels of detail, we also need to enable the study of outliers at the corresponding scales.  
\item \textbf{T3: comparing co-occurrence structures.} 
Common exploration and analysis questions require the comparison of co-occurrence structures that are related to multiple, individual set elements.
Given data about patients with multiple, different diseases, for ex., will easily make it necessary to compare how the individual diseases are associated with symptoms.  
\item \textbf{T4: interactive visual information drill-down.} 
As soon as the analyst reveals relevant information on the overview level, it is highly common that a need for a targeted information drill-down emerges, either to characterize identified patterns (T1), or to explain spotted outliers (T2). Often, the analyst will have to relate to the actual data values, also (detail on demand).  
\end{itemize}
\subsection{Design requirements}
It is usual, as mentioned above, to begin with a univariate analysis, attribute by attribute.  
While this is not the primary focus of CrossSet and existing visualization techniques for set-typed data provide good solutions, already (see Sec.~\ref{sec:related_work}), we still found it necessary to include basic support for such a ``first step'' analysis (\textbf{R1}). 
All other identified design requirements relate to the above described tasks in the bivariate analysis of set-typed data. 
\vspace{0ex}\begin{itemize}\setlength\itemsep{-0.5ex}\raggedright
\item \textbf{R1}: Show the distribution of each of the two set-typed dimensions separately, revealing also the set cardinalities. 
\item \textbf{R2}: Show the counts of individual set elements as well as of the corresponding subsets based on the cardinality for both set-typed attributes at once to reveal co-occurrence patterns (\textbf{T1}). 
\item \textbf{R3}: Enable different levels of aggregation with respect to the set-typed attributes' cardinalities (\textbf{T1}):
(i)~ignore cardinalities, 
(ii)~resolve to a threshold, 
(iii)~cardinalities in full detail.
\item \textbf{R4}: Enable changing the setup of the visual representation of the two set-typed attributes to ease the perception of co-occurrence patterns through complementary perspectives (\textbf{T1}).  
\item \textbf{R5}: Support different visual mappings to highlight data items with many (or few) combinations, support different measures to capture patterns of different types (in particular, rank-based vs.\ value-based patterns), and differentiate combinations with no data items from those with only a few items (\textbf{T2}).
\item \textbf{R6}: Enable complementary views by toggling between a focus on the presence of set elements and one on their absence (\textbf{T3}). 
\item \textbf{R7}: Let the user choose between an element-centric and an item-centric perspective during the analysis (\textbf{T3}).
\item \textbf{R8}:  Show details on demand for selected data-items.
Support a multi-level visualization of set-element distributions for both set-typed attributes (\textbf{T4}): 
(i)~aggregated per individual set-typed dimension, 
(ii)~aggregated for both dimensions at once, 
(iii)~not aggregated, i.e., show all selected individual data items. 
\item \textbf{R9}: Integrate CrossSet into a coordinated multiple views system with linking \& brushing across views. Provide a clear way to indicate brushed data items and, at the same time, provide information on the original distribution (\textbf{T4}). 
\end{itemize}
\noindent 
\subsection{Illustrative dataset}
To illustrate the design of CrossSet, we introduce a \textit{drives} dataset, i.e., a synthetic dataset of more than $250\,000$ car trips, each characterized by two set-typed attributes, called \textit{Input} and \textit{Output}. 
For each trip (\textit{drive}), the \textit{Input}-value is a subset of $\left\{\textit{Music},\,\textit{Family},\,\textit{Traffic},\,\textit{Sport},\,\textit{Aggr}\right\}$, 
depending on whether \textit{music} was played, 
it was a \textit{family} trip, 
there was heavy \textit{traffic}, 
a \textit{sports} car was driven, 
and whether driving was \textit{aggressive}.  
The \textit{Output}-value is a subset of $\left\{\textit{Fun},\,\textit{Resp},\,\textit{Fast},\,\textit{Cheap},\,\textit{Loud}\right\}$, 
depending on whether the trip was \textit{fun}, \textit{responsible}, \textit{fast}, \textit{cheap}, \textit{loud}. 
The \textit{Output}-values conform statistically to a set of co-occurrence rules with respect to the \textit{Input}-values.  
With this approach, we could systematically produce multiple dataset variants to evaluate the suitability of our design for identifying the according patterns.

In the following illustrations of CrossSet (Section~\ref{sec:CrossSet}), we show one of these variants. 
In this \textit{drives} dataset, among the overall $32^2=1024$ possible co-occurrences of \textit{Input}- and \textit{Output}-values, there are, for example, 287 trips where $\left\{\textit{Fun},\,\textit{Resp}\right\}$ (as \textit{Output}-value) co-occurs with $\left\{\textit{Music},\,\textit{Family}\right\}$ (\textit{Input}) and 415 trips where $\left\{\textit{Cheap}\right\}$ co-occurs with $\emptyset$, i.e., the empty \textit{Input}-subset. 
\par 
Central to our work is the study of how values of one set-typed attribute co-occur with the values of another set-typed attribute. In our illustrative \textit{drives} example, we wish to analyze how \textit{Output}-subsets co-occur with \textit{Input}-subsets~-- one example would be: while \textit{Fun} and \textit{Fast} occur comparably often individually, \textit{Fast} co-occurs over-proportionally often with \textit{Aggr}.  
\section{CrossSet design}
\label{sec:CrossSet}

In this section, we describe how our design decisions for CrossSet address the design requirements. CrossSet supports a range of analysis tasks focused on data with two set-typed dimensions. When exploring co-occurrence patterns, analysts may ask: Are certain subsets more frequent than others? or Do specific subsets of one attribute tend to (or tend not to) co-occur with subsets of another attribute? To answer such questions, it is essential to visualize how often different combinations of set-typed attributes appear.

One approach is to represent sets as nodes in a graph, connecting them with weighted edges based on their co-occurrence frequency. However, since the number of subsets grows exponentially, this method becomes impractical even for moderately sized sets.

A more scalable alternative is to adopt a 2D Cartesian layout—an intuitive concept in bivariate data analysis. For example, scatterplots are used for quantitative data and contingency tables for categorical data. CrossSet builds on this foundation by combining aspects of 2D heatmaps and two-way contingency tables. But creating a full contingency matrix for all possible subset combinations is also infeasible. 

Also, any one of the many possible reorderings of such a table would only show some of the co-occurrence patterns due to the ``flattened out'' visual representation~-- while we usually manage to visually compare ``left half'' to ``right half'' in such a table, we generally struggle to visually compare ``every ($4\cdot{}i+3$)\textsuperscript{rd} column'' with ``every ($4\cdot{}i+4$)\textsuperscript{th}''.
To address this problems, aggregation is necessary to manage complexity and reveal meaningful co-occurrence patterns at scale.

\subsection{Basic visual encoding}
We first describe the basic visual encoding of CrossSet, guided by R1 and R2. 
Heatmaps are a natural 2D complement to histograms, visualizing co-occurrences across two data dimensions
(conceptually related to contingency tables in our case). 
Such a heatmap, however, does not scale to all possible subsets per dimension (for any non-small number of set elements).
Instead, we focus on individual set elements, but assign additional rows and\,/\,or columns per element depending on the corresponding set cardinality.

In the context of our illustrative dataset, this means distinguishing cases where only $\left\{\textit{Family}\right\}$ appears as \textit{Input}-value, as compared to $\left\{\textit{Family},\,i_2\right\}$ with $i_2$ being any one of the other four possible set elements, and $\left\{\textit{Family},\,i_2,\,i_3\right\}$, and so on. In this way, we reveal  additional structural information while keeping the heatmap manageable in size.

To explain the basic visual encoding of CrossSet, we first only map one, two, and then three trips with only \textit{Music} and \textit{Family} as set elements of \textit{Input}, as well as \textit{Fun}, \textit{Resp}, and \textit{Loud} as \textit{Output} set elements.  
To start, we consider a trip with \textit{Input}-value $\{\textit{Music}, \textit{Family}\}$ and \textit{Output}-value $\{\textit{Resp}, \textit{Fun}\}$, describing a family trip where music was played, resulting in a fun and responsible drive.
Figure~\ref{fig:basicPrinciple1}:A shows CrossSet of this single data line, including a small heatmap for each combination of set elements. 
Since we have two \textit{Input} set elements and three for \textit{Output}, there are $2\cdot{}3=6$ heatmaps.
Each heatmap spans two rows and three columns, corresponding to the possible set sizes (up to 2 for \textit{Input} and up to 3 for \textit{Output}).
The considered drive is accounted for in $2\cdot{}2=4$ heatmaps: \textit{Music}\,/\,\textit{Fun}, \textit{Music}\,/\,\textit{Resp}, \textit{Family}\,/\,\textit{Fun}, and \textit{Family}\,/\,\textit{Resp}. In each of them, it's accounted for in the ``$+1$\,/\,$+1$'' cell, corresponding to ``$\textit{Music}+1$\,/\,$\textit{Fun}+1$'', ``$\textit{Music}+1$\,/\,$\textit{Resp}+1$'', etc.~ 
The drive belongs to four bins, and we add it to the heatmap with a value of one-quarter in each of the four bins.
The marginal histograms on the top and on the right show the corresponding \textit{Input}- and \textit{Output}-sums.  
\par 
Next, we consider a second trip in addition, where $\{\textit{Traffic}\}$ (\textit{Input}) co-occurs with $\{\textit{Resp}\}$ (\textit{Output}), shown with CrossSet in Figure~\ref{fig:basicPrinciple1}:B.
It's accounted for in heatmap \textit{Traffic}\,/\,\textit{Resp} (in cell ``$\textit{Traffic}$\,/\,$\textit{Resp}$'').

We then add a third trip where \textit{Input}-value $\{\textit{Traffic}\}$ co-occurrs with \textit{Output}-value $\{\textit{Resp}, \textit{Fun}\}$. 
Since the set sizes of the set-typed attribute values are 1 and 2, this additional drive is accounted for in two heatmaps: \textit{Traffic}\,/\,\textit{Fun} and \textit{Traffic}\,/\,\textit{Resp} (in cells ``$\textit{Traffic}$\,/\,$\textit{Fun}+1$'' and ``$\textit{Traffic}$\,/\,$\textit{Resp}+1$''), as shown in Figure~\ref{fig:basicPrinciple1}:C. 
\par 
The marginal histograms provide a 1D distributional overview with respect to the different subset structures per attribute (R1).  
They account for all contributions to the rows and columns, and their total sum equals the number of data items. 
The sums are expressed as fractions to reflect the fact that all contributions add up to 1 per item. 
The numerator represents the number of drives, while the denominator corresponds to the set cardinalities. 
Fraction $\sfrac{2}{2}$ in Fig.~\ref{fig:basicPrinciple1}, for example, means that 2 drives are accounted for, but since they show up in two histogram bins, their contribution is split, accordingly.
\par
\subsection{Analyzing co-occurrence patterns}
\begin{figure}[t]
\centering
\includegraphics[width=.45\linewidth]{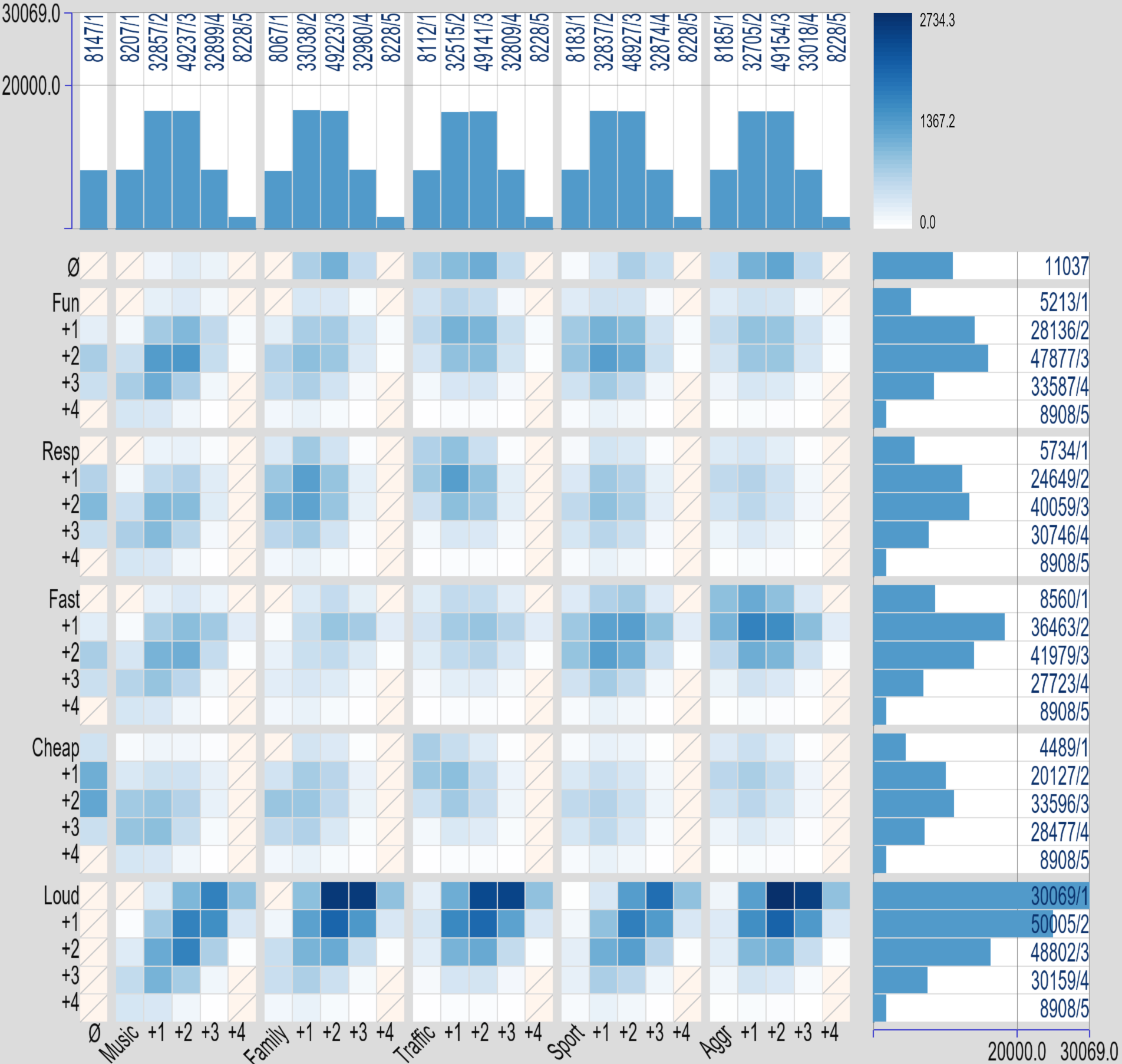}
\includegraphics[width=.45\linewidth]{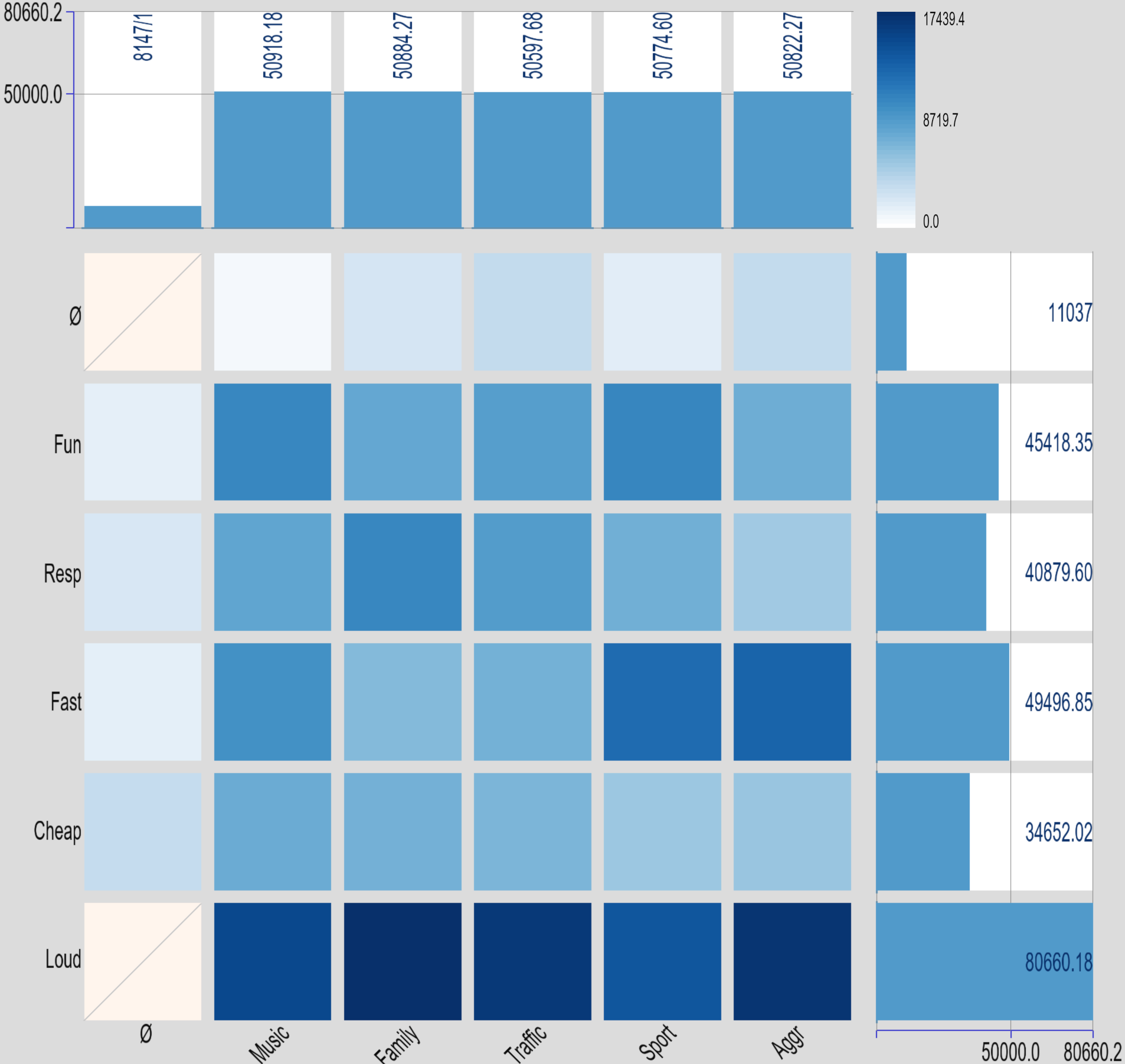}
\caption{CrossSet with fully expanded heatmaps on the left and fully collapsed ones (one bin per element--element co-occurrence) on the right supports the exploration of co-occurrence patterns at different levels.
}
\label{fig:allCollapsed}
\end{figure}

The basic visual encoding of CrossSet already supports pattern detection (R1 and R2) and outlier analysis to a certain degree (R5). 
Figure~\ref{fig:allCollapsed}, left and Figure~\ref{fig:drive_all_1}  (suppl.\ material) show the full drives dataset visualized with CrossSet. The marginal histograms show distribution information regarding the individual attributes. 
While the \textit{Input}-values (differentiated horizontally) have similar individual occurrences (every possible \textit{Input}-subset occurs approximately equally often), \textit{Output}-values including \textit{Loud} are particularly prevalent.  
The heatmap showing \textit{Output}-subsets including \textit{Loud} vs.\ all sorts of \textit{Input}-values (lowest row of heatmaps) show that \textit{Loud} comes often alone, or plus one more other set element. 
When \textit{Loud} occurs alone, the \textit{Input}-subsets are mostly of size 3 or 4 (``+2'' and ``+3'' columns)~-- roughly judging: the more \textit{Input} set elements are present, the fewer \textit{Output} set elements (given that \textit{Loud} is included), and vice versa. 
When comparing ``\textit{Loud} heatmaps'' with ``\textit{Cheap} heatmaps'' (row above), we see a different overall pattern: the fewer \textit{Input} set elements are present, the fewer \textit{Output} set elements, also (given that \textit{Cheap} is included), and vice versa. It's notable, however, that this ``positive correlation'' pattern only shows up for subset sizes up to 2 or 3.  
Such co-occurrence patterns would be impossible to see, when only visualizing the set-typed attributes individually.  
\par
To also account for empty subsets ($\emptyset$) as valid values of set-typed data dimensions, we represent them as ``special column\,/\,row'' in CrossSet on the very left\,/\,top.  
Interestingly, the visualization of empty sets is often neglected in the visualization of set-typed data. 
The top-right marginal count shows that there are 11\,037 trips with $\emptyset$ as \textit{Output}-value. None of them has only \textit{Music} or \textit{Family} as \textit{Input}-value, and none has all five \textit{Input} set elements.  
Also, to clearly differentiate empty bins (from bins with a low count) in the heatmaps, we use a dedicated representation for empty bins (see the light yellow, crossed out cells in Fig.~\ref{fig:drive_all_1}). 
\subsubsection{Collapsing and uncollapsing of levels} 
To support pattern detection at different scales of aggregation (R3) we enable the flexible aggregation of selected rows and\,/\,or columns.
Up to now all cardinalites were shown (R3(iii)).
\par
Heatmap rows\,/\,columns can also be fully collapsed individually, depicting them with a single row or column (R3(i)). 
Figure~\ref{fig:allCollapsed}, right shows the drives data in CrossSet with all heatmaps fully collapsed. Such a configuration, showing high-level overall co-occurrence patterns, serves as a useful starting point for the analysis.
The labels of the columns and rows reflect such a parameterization. 
In Fig.~\ref{fig:synthCS}, for example, three of the four heatmap rows (for $b_2$, $b_3$, and $b_4$) and two of the heatmap columns ($a_3$, $a_4$) are collapsed. 
Figure~\ref{fig:collapsingExamples} (suppl.\ material) shows some other configurations that illustrate level reduction by collapsing individual set elements.

CrossSet can also be parameterized to show no more than a certain number of rows\,/\,columns per heatmap~-- this is important, when set-typed attributes are visualized with a large number of set elements (R3(ii)).  
If the cardinalities are limited to no more than $2$, for example, the last row\,/\,column includes all larger subset sizes, as well. The label then changes from ``$+2$'' to ``$+2\ldots$''. In addition, marginal histograms which show fractions (in case of item-centric visualization) do not show fractions but decimal numbers instead as several fractions are merged to a single row or column. Such details are needed to clearly communicate what is shown.
Figure~\ref{fig:partiallyCollapsed} (suppl.\ material) shows the drives data set with reduced cardinality levels.

\subsubsection{Heatmap reordering} 
Reordering of rows and\,/\,or columns in heatmaps (and other tabular visualization methods for unordered data) is often used to facilitate the detection of relevant patterns~\cite{Behrisch_Matrix_2016}. To support changing the CrossSet perspective accordingly (R4), we provide flexible means to change the order of the set elements. Especially, when certain set elements from dimension \#1 correspond to certain set elements from dimension \#2 in a semantically relevant way, then ``aligning'' them in the respective ordering of rows and columns helps with the analysis.  We demonstrate this effect in Fig.~\ref{fig:reorder_comparison} (suppl.\ material).  

\subsection{Studying outliers}

As previously described, we highlight empty bins using distinct color coding and an additional crossing line to clearly distinguish them from bins containing only a few items (outliers) (R5). Support for the remaining requirements from R5 in CrossSet is described as follows.

\subsubsection{Value-based vs.\ rank-based visualization} 
If the differences between the cell counts are large, a linear color map may struggle to resolve various details. 
To support an analysis across different scales we provide rank-based coloring as an alternative, i.e., coloring the rank of the cell counts, instead of their actual value. 
This mode makes better use of the available color contrast, assisting comparative analysis questions that do not depend on actual counts. 
In the case of ties, i.e., cell counts with the same value, we provide standard competition ranking and dense competition ranking~\cite{Papoulis91}. 
The tooltip shows which mode is active. 
Figure~\ref{fig:mappings}:B (suppl.\ material) shows the rank based mapping for all drives. 
\subsubsection{Showing deviations from expected values} 
With CrossSet, we can display deviations from certain expected values. By default, we assume an even distribution of all possible subsets. Accordingly, we provide a mode, where the relative proportion to this default assumption is showing (instead of the actual counts).

For this mode we need a divergent color scale.
Here, it is important to normalize the scale so that the minimum and the maximum value correspond to the same proportional deviation. If the maximum deviation is much larger on one end, we need to scale the other end accordingly.  

\begin{figure}[t]
\centering
\includegraphics[width=\linewidth]{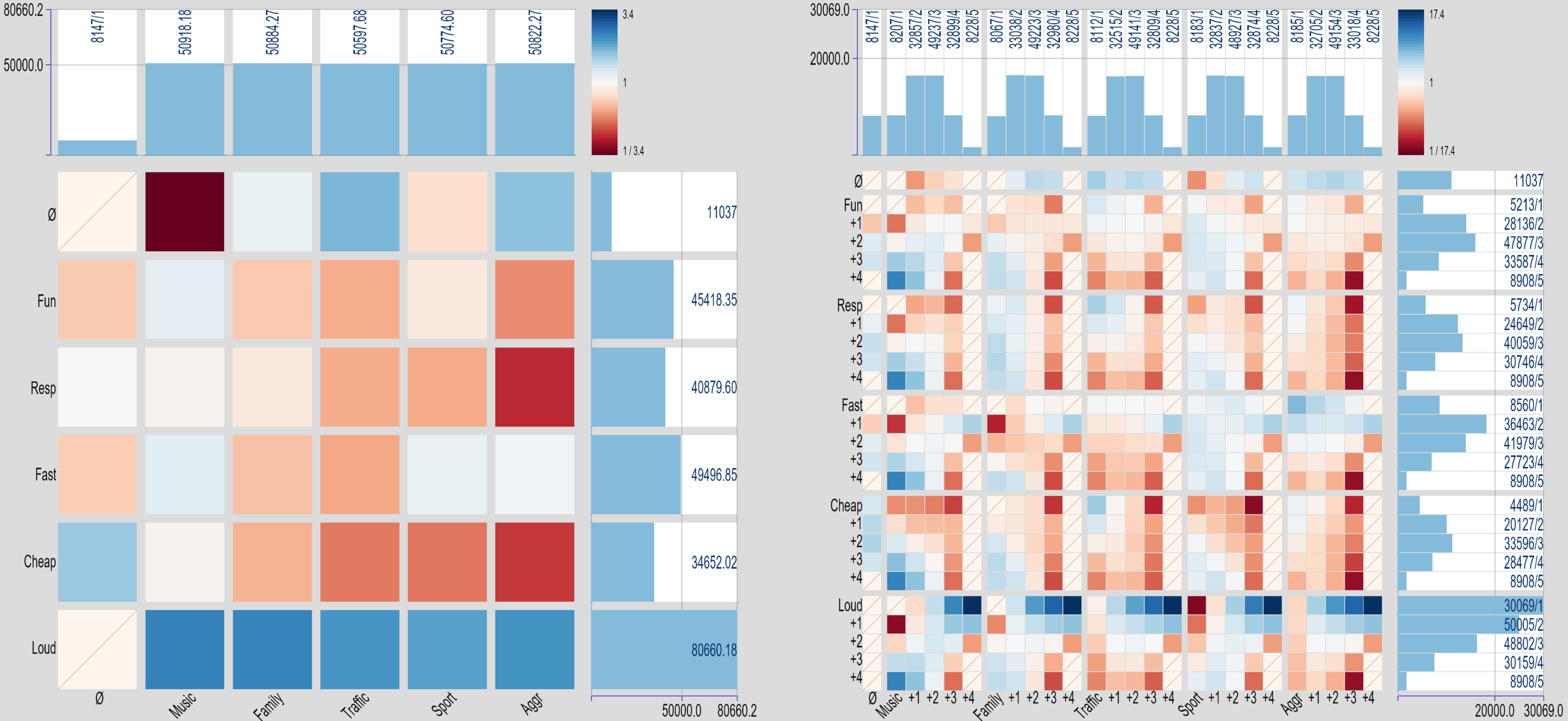}
\caption{A diverging color scale shows deviations from expected values. Here, the deviation from an even distribution is shown for collapsed and non-collapsed CrossSet. The blue bins indicate over-represented counts.}
\label{fig:normalized}
\end{figure}

This mode enables a quick understanding of how set elements and their combinations are over- or underrepresented in the data. Figure~\ref{fig:normalized} shows deviations for both the fully collapsed and for the uncollapsed CrossSet. It is evident from the collapsed CrossSet figure that {\textit{Loud}} is overrepresented, while {\textit{Aggr+Cheap}} as well as {\textit{Aggr+Resp}}, for example, are underrepresented.

\subsection{Comparing co-occurrence structures}
Requirements R6 and R7 originate from the task of comparing co-occurrence structures.
This section describes how these requirements are fulfilled in CrossSet.

\subsubsection{Counting the presence or absence of elements}
In the analysis of Boolean attributes it is often natural to invert a variable. The same holds for set elements. We can be interested in patients with fever, or without. Inverting an individual set-element makes it possible to focus on analysis questions with respect to the presence or absence of a set element (R6). Any such negation is represented in the label (``L'' vs.\ ``$\neg$L'').  As CrossSet represents aggregated set cardinalities, inverting an element leads to an alternative perspective~-- instead of counting {\textit{Loud}} drives, for ex., \textit{silent} drives are counted. 
Further, in our example, all outputs except for \textit{Loud}, are better if present more prominently, so inverting \textit{Loud} aligns the set elements semantically.  
Figure~\ref{fig:silent} (suppl.\ material) shows the data with inverted {\textit{Loud}} perspective. 
Note the differences to the original CrossSet configuration, shown in Figure~\ref{fig:drive_all_1} with many ``{\textit{Loud}} only'' drives~-- in the $\neg$\textit{Loud} perspective, they show up as $\emptyset$ (as \textit{Output}-value). 
\subsubsection{Counting data items vs.\ set elements} 
In the basic mode, as described above, we count each data item once. 
If it belongs to multiple bins, only an according fraction is accounted for. 
The sum of the marginal histograms equals the number of data items.

If we are interested in how many drives were \textit{Fun}, for example, we count each contribution to a bin as one. 
This way, we show how often each set element is present in the data.
The user can easily switch between these two modes (R7).
The difference compared to the basic mode is more pronounced for bins showing sets of higher cardinality, as these drives are counted once per set element (and not just once overall).
As a consequence, the total sum of marginal histograms amounts to the number of set elements (usually larger than the number of data items).
Figure~\ref{fig:mappings}:A (suppl.\ material) shows the set elements count mode of CrossSet for the drives data. The counts in the marginal histograms are shown as integers now to clearly indicate counting of set elements. 

Depending on the data and on the counting mode, the distribution of counts can differ a lot. 
The color mapping can be used to emphasize differences among larger or smaller values.
As the counts distribution is not known in advance, we support several color maps which are nonlinearly mapped between white and the darkest color.  
\subsection{Interactive visual information drill-down}
\begin{figure}[t]
\centering
\includegraphics[width=\linewidth]{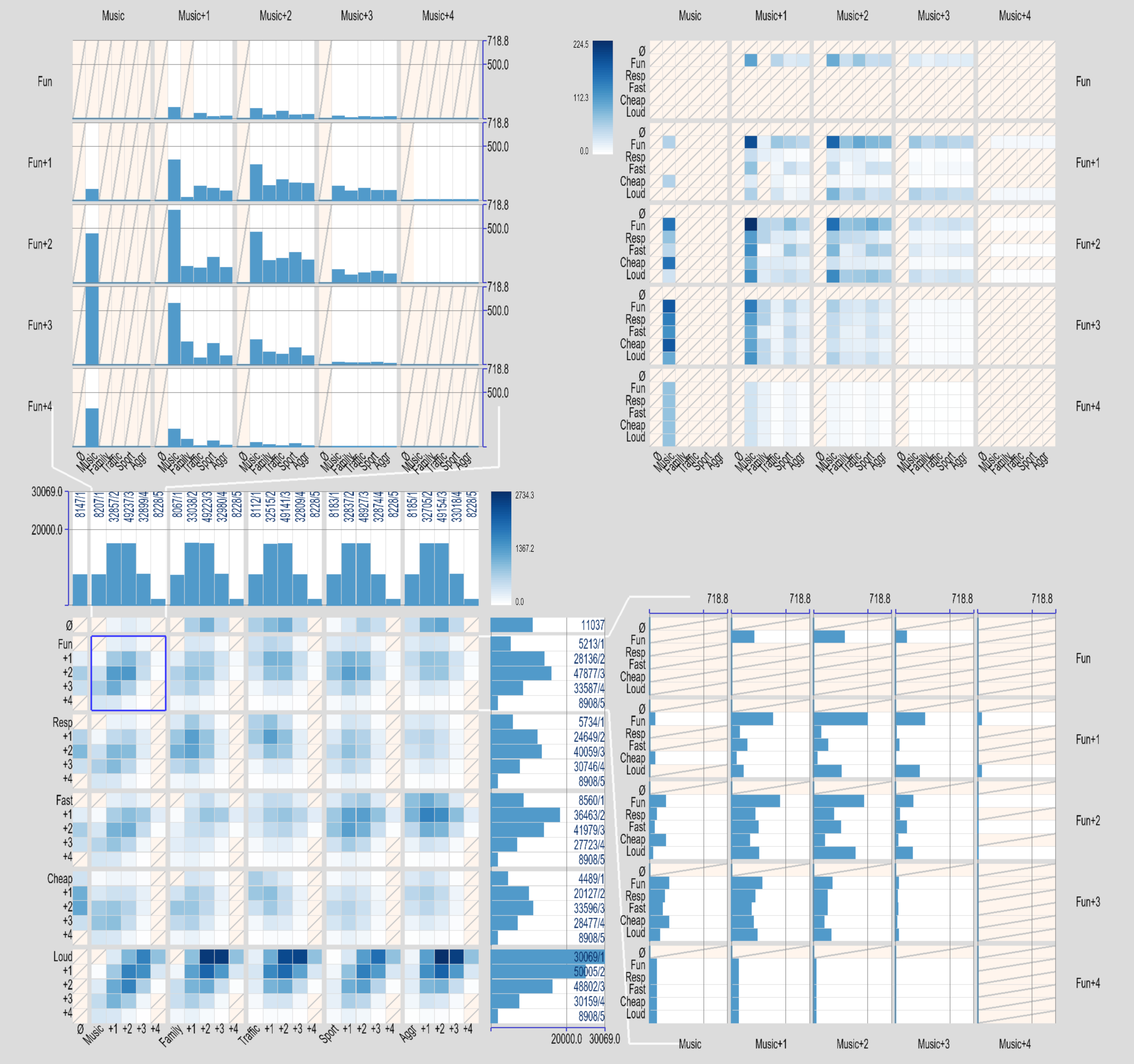}
\caption{Detail histograms (top-left and bottom-right) and a detail heatmap (top-right) provide more detailed counts. For the selected heatmap in the lower-left (blue rectangle), one detail histogram is shown for each heatmap cell, providing insight into what the aggregations ``+1'', ``+2'', etc., stand for in the heatmap. 
The detail heatmap (top-right) relates these histograms to each other.
}
\label{fig:allDetails_First_small}
\end{figure}
\begin{figure}[t]
\centering
\includegraphics[width=\linewidth]{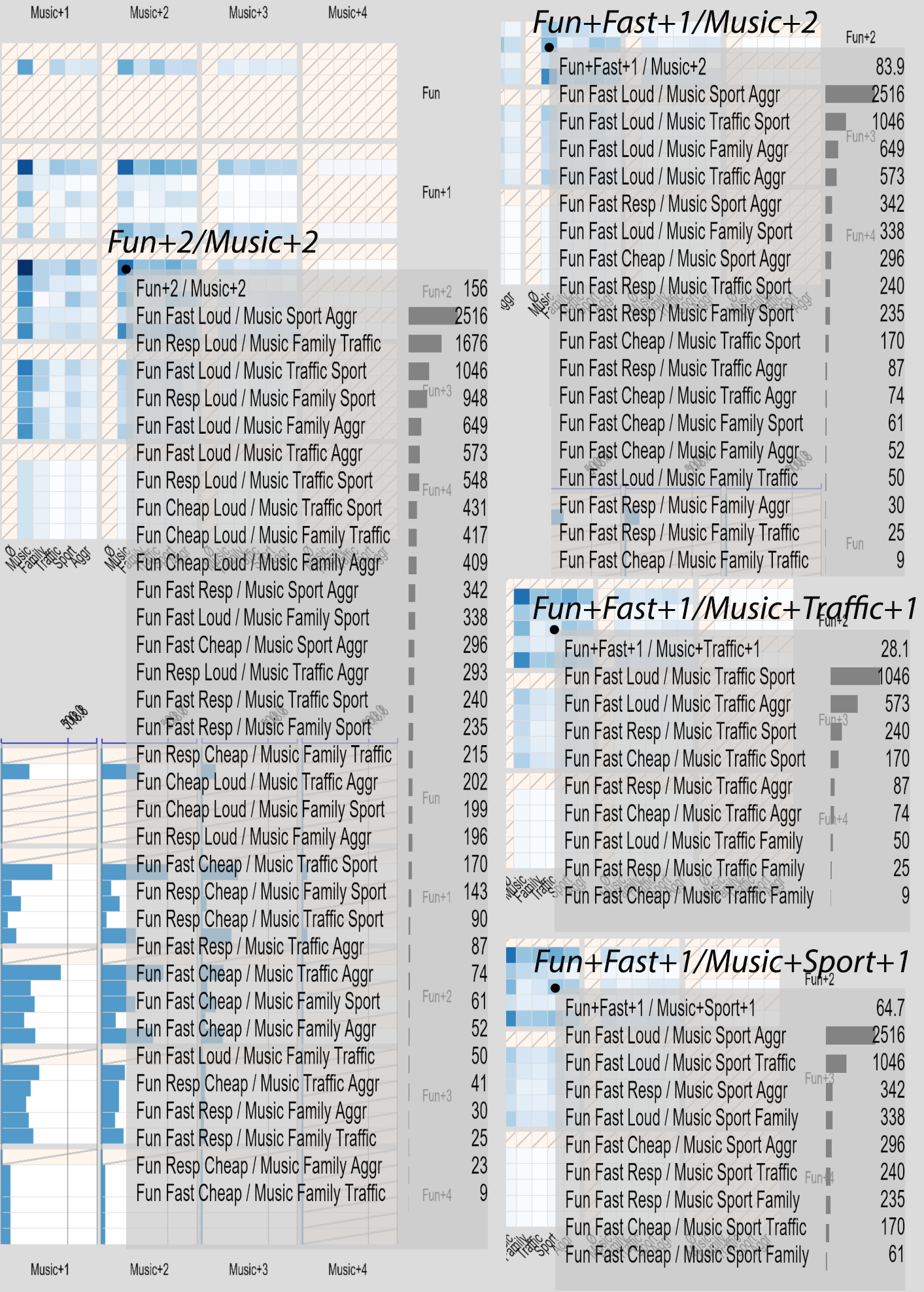}
\caption{For full details on demand, a mouse-over tool tip reveals the counts of all co-occurrences of the per-attribute subsets (for the detail heatmap cell under the mouse cursor), ranked by their number.}
\label{fig:tooltips_001}
\end{figure}
Except for unnaturally small datasets, we depend on aggregation (at different levels) to convey a meaningful visualization. 
In Figure~\ref{fig:drive_all_1} (suppl.\ material), for example, we see that 
$\textit{Fun}+2$ co-occurs prominently with $\textit{Music}+1$ as well as with $\textit{Music}+2$ (top-left heatmap).
At this level of aggregation, however, we do not see which of the other \textit{Input} and \textit{Output} set elements stand in for ``+1'' and ``+2''.

Showing all possible combinations would require a lot of space and high-level details would be lost.
To still support a more detailed analysis (R8), we extend the basic encoding and make details accessible on demand through carefully designed interaction. 

Selecting a heatmap reveals additional details related to the selected row and column (R8(i)). 
Figure~\ref{fig:allDetails_First_small} shows an example with more details for the {\textit{Music}} vs.\ {\textit{Fun}} heatmap. 
The selected heatmap has 25 cells. 
The detail histograms on the top of the original CrossSet view show the 25 histograms with distributions of the \textit{Input} set elements for each of the cells of the heatmap. 
The detail histograms on the right show the distributions of the \textit{Output} set elements for the same cells.  

Examining the detail histograms for the {\textit{Fun+1}}, {\textit{Fun+2}}, and {\textit{Fun+3}} rows (top-left), we see that in the case of {\textit{Fun+1}}, {\textit{Family}} co-occurs rather rarely (together with \textit{Music}), while {\textit{Traffic}}, {\textit{Sport}}, and {\textit{Aggr}} co-occurs more often.  
In the case of {\textit{Fun+2}}, {\textit{Family}}
does not distinguish itself in this way, and 
in the case of {\textit{Fun+3}}, {\textit{Family}} even co-occurs rather often (together with \textit{Music}, similar to \textit{Sport}).
Considering the histograms on the right, we see how \textit{Output} set elements are distributed along with {\textit{Fun}}~-- all in the case of driving with \textit{Music}.
In several histograms we see two counts that are larger than the others (besides {\textit{Fun}}, which is always present, by definition), corresponding to additional \textit{Output} set elements {\textit{Fast}} and {\textit{Loud}}. 

Even more detailed information is provided by 
the detail heatmaps (with 25 individual heatmaps) in the top-right, providing distributions of all combinations for each original cell, and consequently, for each of the detail histograms (R8(ii)).

Considering the {\textit{Fun}+2} vs.\ {\textit{Music}+2} heatmap, i.e., the central detail heatmap in the top-right, we see horizontal ``stripes'' that correspond to the larger counts in the lower-right detail histograms: {\textit{Fun}+2} comes often with {\textit{Fast}} and {\textit{Loud}}, but in this heatmap we also see distributions across \textit{Input} set elements. 
{\textit{Fast}} is co-occurring less with {\textit{Family}} and {\textit{Traffic}} while driving with {\textit{Music}}. 

Even though we have drilled into more details, already, we still need one more level of detail to see all possible combinations (R8(iii)). 
On mouse-over, we thus provide a detail tool tip, which shows the number of individual combinations in the corresponding bin (see
Figure~\ref{fig:tooltips_001}).
The detail tool tip shows all contributing combinations for the heatmap cell which the mouse hovers over, for example {\textit{Fun}+2} vs.\ {\textit{Music}+2}. 
We show the total count of the cell, taking into account that not every drive contributes with 1, depending on the heatmap and the cell. 
We also show the ``aggregation rule'' for the heatmap cell, and then all individual combinations, ranked by the number of accounted data items (inspired by UpSet~\cite{UpSet_2014}), and showing the actual items count. While this may result in lengthy tooltips, we chose to display the full list to ensure data completeness and avoid adding further complexity to the interface.

Moving the mouse cursor to the cell right below, staying in the  {\textit{Music}+2} column, we resolve ``+2'' (along with \textit{Music}) with ``\textit{Fast}+1''. 
The tool tip shows {\textit{Fun+Fast+1}} vs.\ {\textit{Music+2}}, accordingly. 

Moving the cursor to the right, we narrow {\textit{Music}+2} down to {\textit{Music+Traffic}+1}, 
with the tool tip showing all combinations for {\textit{Fun+Fast+1}} vs.\ {\textit{Music+Traffic+1}}.
Most often, the following subsets co-occur: 
{\textit{Fun+Fast+Loud}} and {\textit{Music+Traffic+Sport}} (1046 drives) and {\textit{Fun+Fast+Loud}} and {\textit{Music+Traffic+Aggr}} (573).
In the case of {\textit{Fun+Fast+1}} vs.\ {\textit{Music+Sport+1}}, the most common co-occurrence is:
{\textit{Fun+Fast+Loud}} and {\textit{Music+Sport+Aggr}} (2516 drives).
\begin{figure}[t]
\centering
\includegraphics[width=\linewidth]{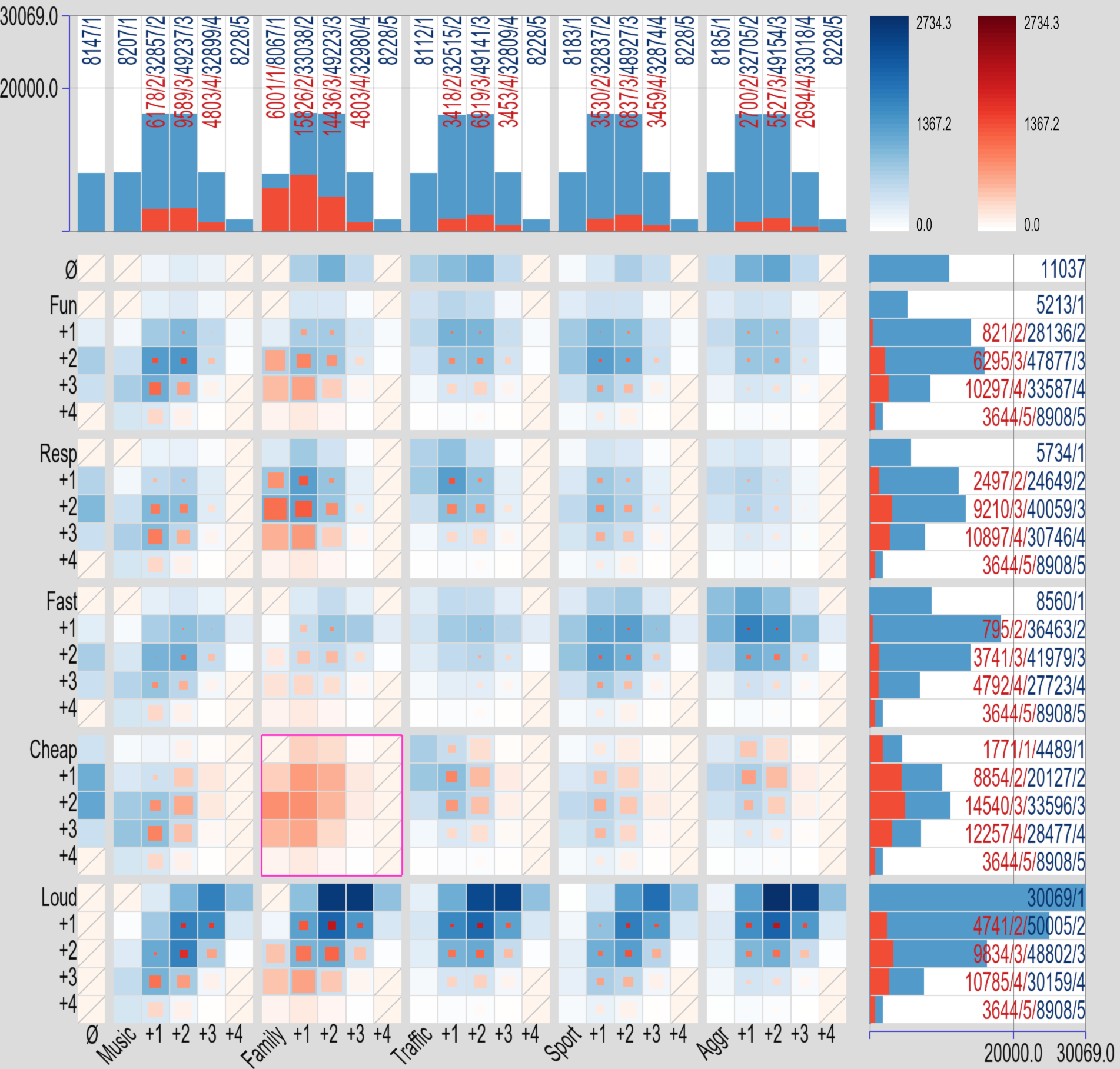}
\caption{In CrossSet, any heatmap, any histogram, or any bin can be brushed (selected), and all corresponding
data items in all other bins are highlighted. Here the co-occurrence of {\textit{Family}} and {\textit{Cheap}} is selected. The brushed bins are colored using a different color scale with the same lightness mapping. The lightness indicates the original value, and the number of brushed items in a bin is shown by the size of the red bin part.} 
\label{fig:brush_example_001}
\end{figure}

\par 
CrossSet is fully integrated with a coordinated multiple views system and linking\,\&\,brushing is supported across all views (R9). 
Figure~\ref{fig:CMV} (suppl.\ material) shows CrossSet together with other views in the coordinated multiple views system. The CrossSet control pane is also shown in the figure.
Any heatmap, any histogram, any heatmap cell, and any histogram bin can be brushed
in CrossSet. Composite brushes are also supported, i.e., selections that are combined using Boolean operations.
We use two related data scales to show brushed data in CrossSet heatmaps so that the distribution of the original cell is still encoded in a fully brushed cell.
Hence the brushed data color corresponds to the shade of the original color.
Figure~\ref{fig:brush_example_001} shows an example where the user brushed the {\textit{Family}} vs.\ {\textit{Cheap}} heatmap.
The heatmap itself is highlighted with a bounding rectangle, and the brushed bins are shown using a red color scale.
All bins in the selected heatmap are brushed. The corresponding drives in other heatmaps and in the marginal histograms are highlighted, accordingly.
The number of brushed drives is indicated by the size of the red bins and the color coding of the red bin corresponds to the original color coding.

In this example, we observe interesting patterns. 
The marginal histogram for {\textit{Music}} show more brushed drives (including {\textit{Family}} and {\textit{Cheap}} according to the selection) than those for {\textit{Traffic}}, {\textit{Sport}}, and {\textit{Aggr}}. The difference is larger for +1 cases. Cheap family drives co-occur comparably rarely with heavy traffic, with a sports car, or with aggressive driving.
Figure~\ref{fig:colorScales} (suppl.\ material) shows the blue and the red color scales, distributed to support emphasizing variations among larger and smaller values. Per default, the neutral (linear) scale is used.
\section{Demonstration}
\label{sec:evaluation} 
We analyze two datasets to demonstrate the effectiveness of CrossSet:
A movies dataset~\cite{movies-data} and a patients data set.
These datasets have different characteristics and their analysis demonstrates different strengths of CrossSet.
In the  patients data set the two set-typed dimensions draw from (roughly) the same number of set elements each, while the situation is unbalanced for the movies.
The demonstration of the patients dataset can be found in the supplementary material, Section~\ref{suppl:additinalCase}.
%
%
%
\begin{figure}[t!]
    \centering
    \begin{subfigure}[t]{0.24\textwidth}
        \centering
        \includegraphics[width=\textwidth]{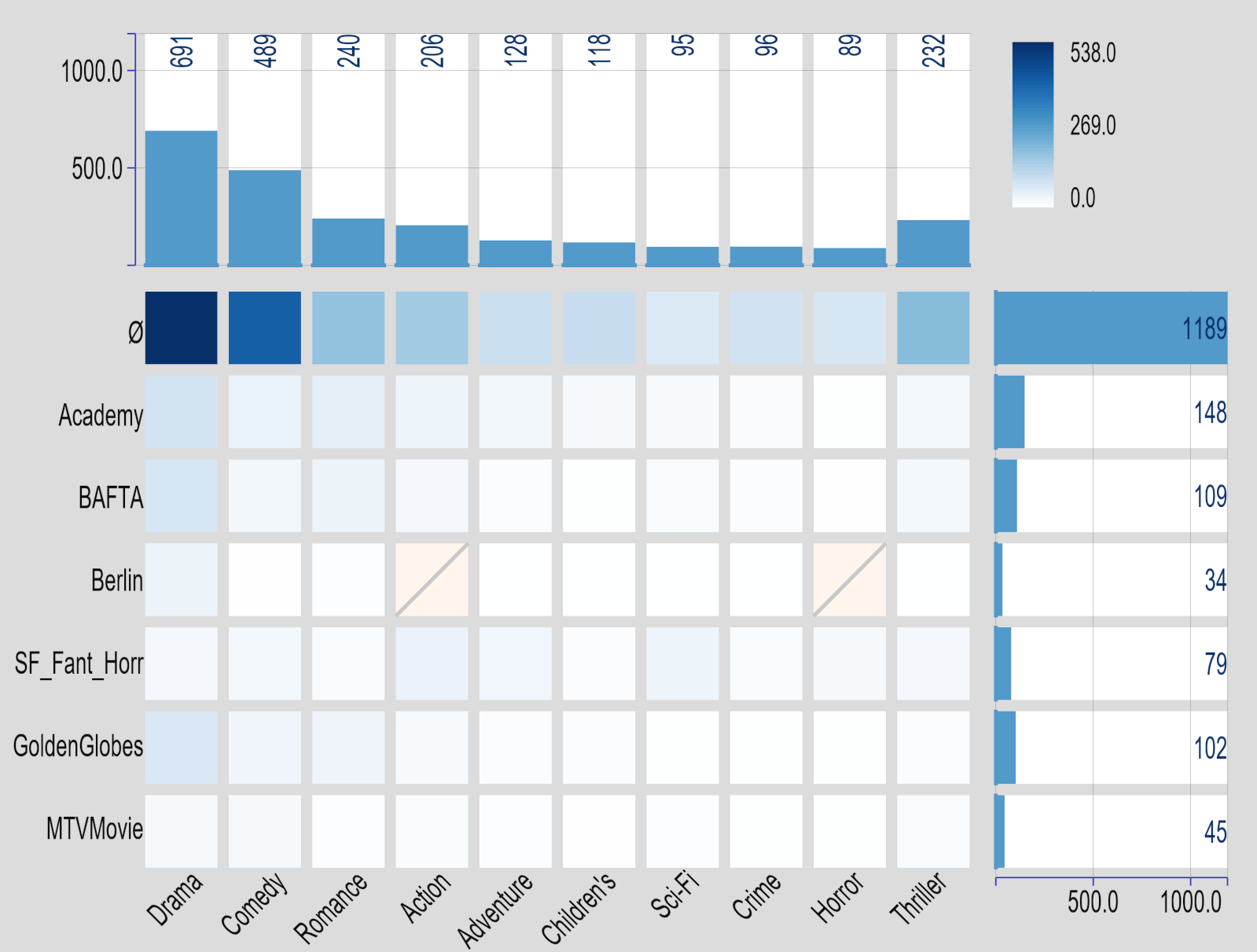}
           
    \end{subfigure}%
    ~ 
    \begin{subfigure}[t]{0.24\textwidth}
        \centering
        \includegraphics[width=\textwidth]{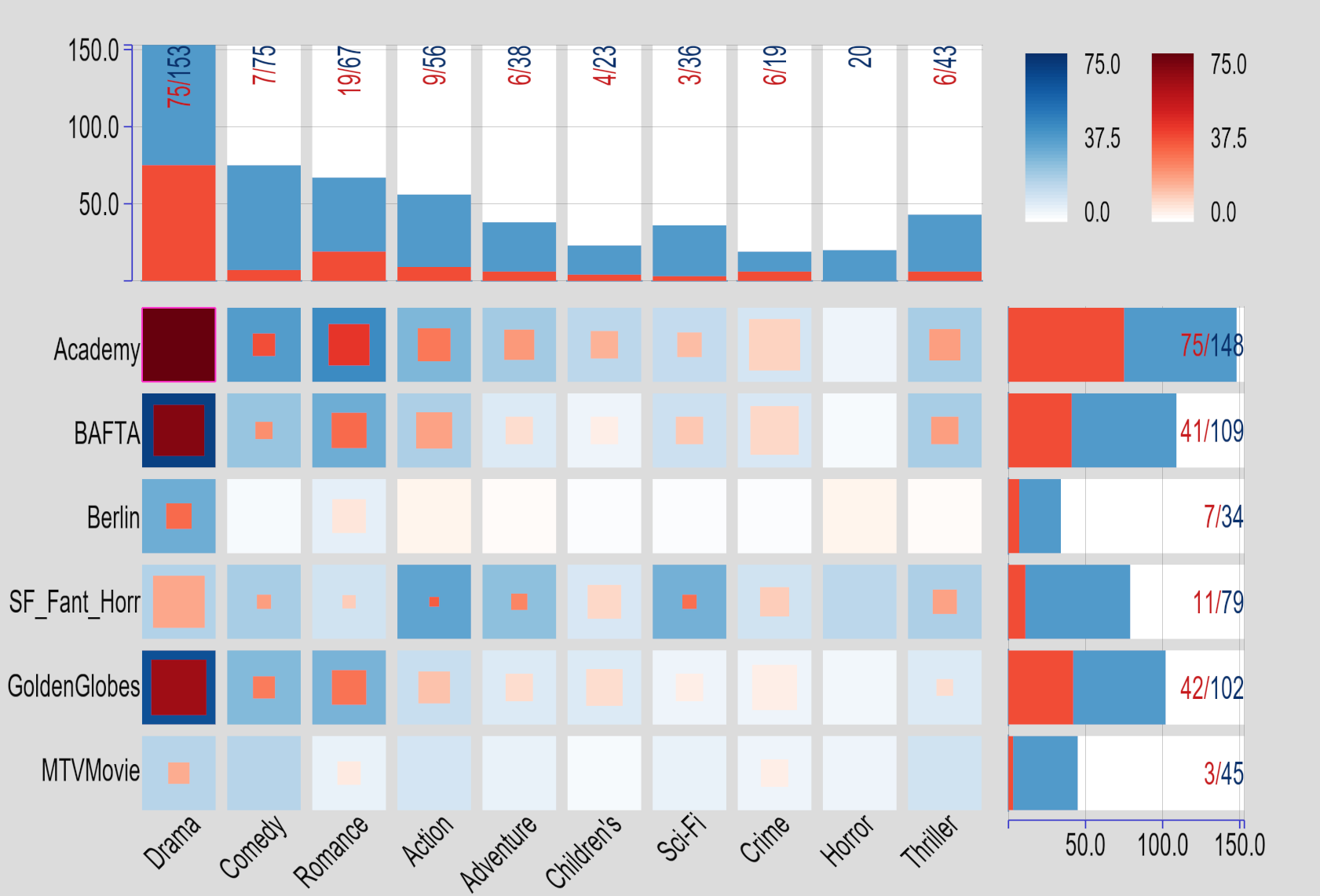}
        \end{subfigure}

    \caption{Overview for movie data set by using fully collapsed set elements with \textbf{(left)} and without \textbf{(right)} empty sets for awards. \textbf{(left)} The fully collapsed heatmap for the Movies dataset. Many movies have not received an award, and only two cells are empty (we highlight them with line over and yellowish color). \textbf{(right)} Empty sets are hidden to better see the patterns for movies that have won an award. 
One cell with the highest count is brushed, \textit{Academy} / \textit{Drama}.
\textit{BAFTA} and \textit{Golden Globe} awards are strongly linked with the \textit{Academy} award for this genre.
The selected genre \textit{Drama} co-occurs most often with the genre \textit{Romance} when winning an \textit{Academy} award.}
\label{fig:movies_overview} 
\end{figure}
\par 
The drives dataset is characterized by a limited number of set elements per set-typed attribute. 
Clearly, we may have data with more set elements and fewer combinations. 
With a na{\"{i}}ve approach, such data would make CrossSet large and sparse. 
The ability to collapse rows and columns, however, and to limit the maximally shown cardinality helps in such cases. 
\par 
The movies dataset~\cite{movies-data} is such a case.
It has been often used to demonstrate visualization and analysis methods for set-typed data~\cite{Alsallakh-2013-a,UpSet_2014,Matkovic-2020-c}. 
In all of these earlier works sets of genres were analyzed.
Here, we extended the data with another set-typed dimension~-- awards. 
Each movie has a set of genres and a set of awards.
Besides two set-typed dimensions, $\textit{Awards}$
and $\textit{Genres}$,
the data also includes rating and year for all 1521 movies. 
This dataset lends itself to a bivariate analysis, where we are interested in the interaction between movie genres and awards, e.g., we can ask a question like: Is there is a winning combination of genres for a specific combination of awards?
Are some genre combinations more likely to win?  
Chung et al.~\cite{GridSet_2020} asked similar questions. 
They have one set-typed attribute, genres, and they analyze which combination of genres is more common for one particular award. 
In our case, awards are also of set-type, so we can analyze how combinations of awards relate to combinations of genres. 
This is the key difference: We focus on the bivariate analysis of two set-typed dimensions.
\par 
Although empty sets often represent important cases, they are rarely  shown in set visualization methods explicitly.
We allow them to be shown or hidden, depending on tasks and data characteristics.
In this case, we hide the empty sets for genres because of the data characteristics, i.e., at least one genre is defined for all movies, and we show or hide the empty sets for awards to support the task at hand.

Figure~\ref{fig:movies_overview} (left) shows a CrossSet visualization  where all set elements are collapsed to get a high-level aggregated overview including the empty sets for awards. 
We can clearly see that most of the movies have not won any award.
Still, almost all heatmap cells are populated, i.e., movies of all different genres have won almost all awards.
Just the \textit{Berlin} film festival award was not given to any action or horror movie.
The cells as shown in Figure~\ref{fig:movies_overview} (left) include all available combinations of the elements, e.g., the cell \textit{Drama} contains all movies with \textit{Drama} as one of its genres.
\par 
To better see the patterns for movies that have won an award, we hide the empty sets for awards. 
Figure~\ref{fig:movies_overview} (right) still shows the overview level but now without the empty set and one brushed cell, 
\mbox{\textit{Academy}\,/\,\textit{Drama}}, the cell with the highest count.
We can see that the \textit{BAFTA} and \textit{Golden Globe} awards are strongly linked with the \textit{Academy} award for \textit{dramas}.
Since we still show the overview level, all combinations with other genres are also highlighted.
The selected genre \textit{Drama} co-occurs most often with the genre \textit{Romance} when winning an \textit{Academy} award.
Again the \textit{BAFTA} and the \textit{Golden Globe} awards show strong similarities to this pattern.
Most of the \textit{SF} awards for the genre \textit{Drama} are also highlighted but the combination with \textit{Romance} is rare.
\begin{figure}[t!] 
\centering
\includegraphics[width=\linewidth]{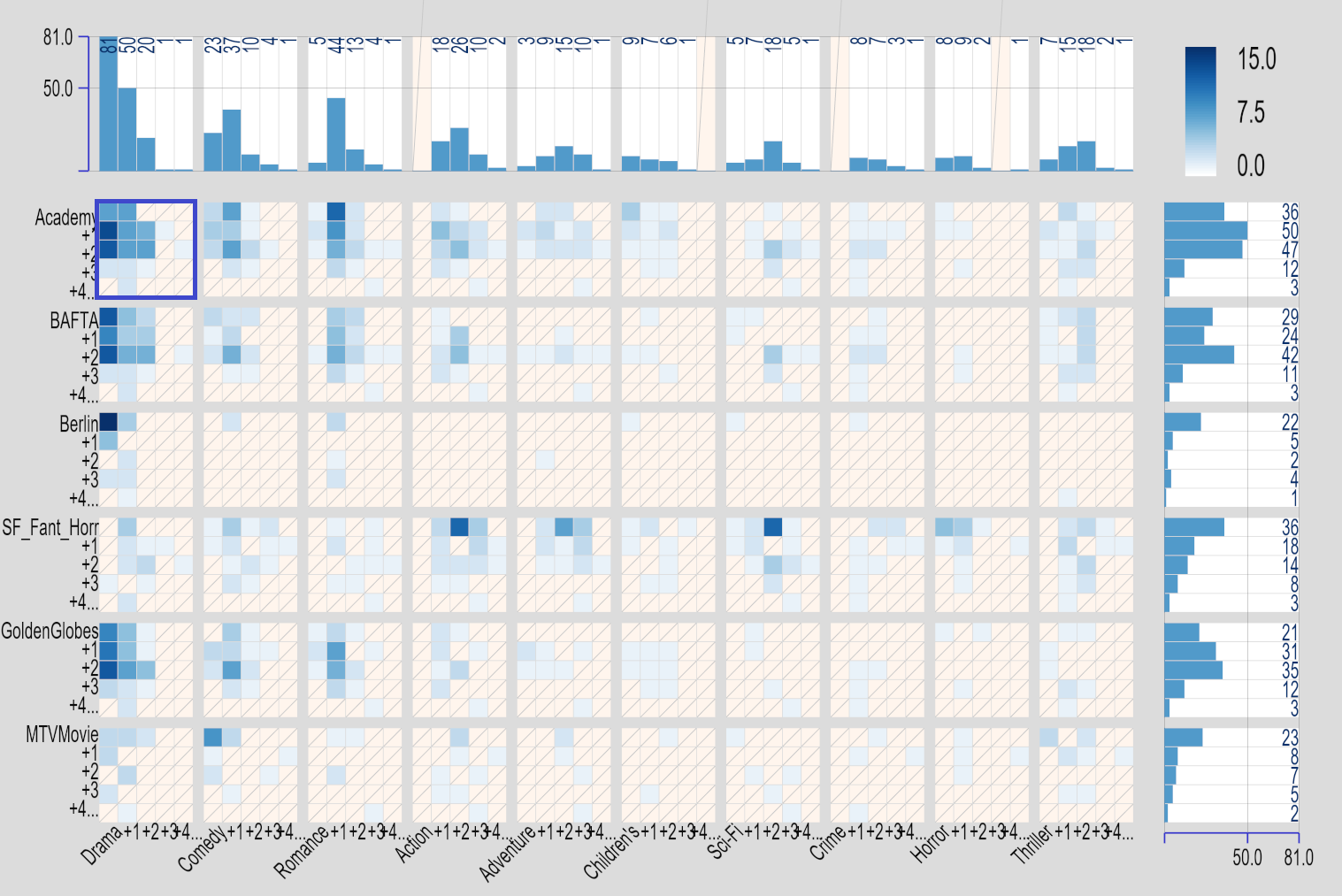}\\[-1.5ex] 
\caption{
Expand set elements to cardinality five to show patterns in the heatmap and the marginal histograms. Note the strong co-coccurance patterns for 
\textit{Academy}, \textit{BAFTA} and \textit{GoldenGlobes} especially for genres \textit{Drama}, \textit{Romance} and \textit{Comedy}. }
\label{fig:movies_analysis_patterns}
\end{figure}
\par
We now expand both set-type dimensions to enable further in-depth bivariate analysis. 
Having ten genres and six awards leads to a rather extended heatmap. 
Since there are no movies with many genres or awards, we limit the cardinality of genres and awards to five. 
The +4\ldots{} columns and rows contain movies with five and more genres\,/\,awards. 
Figure~\ref{fig:movies_analysis_patterns} shows CrossSet in this configuration. 
The marginal histograms confirm that the limits are reasonable.
If the limits would be too low, the last bin in the marginal histograms would be dominant. 
The marginal histograms for genres, on the top of Figure~\ref{fig:movies_analysis_patterns}, indicates that \textit{drama} is the most frequent genre and most often does not co-occur with others. 
Almost all other genres most often co-occur with others. 

When looking at the marginal histogram for awards, on the right of Figure~\ref{fig:movies_analysis_patterns}, the \textit{MTV} award and the \textit{Berlin} award do not co-occur very often with other awards.
On the other hand, \textit{Academy}, \textit{BAFTA} and \textit{GoldenGlobes} awards co-occur very often with two other awards.
When switching our focus to the heatmap we see that the patterns in the heatmap cells for the \textit{Academy}, \textit{BAFTA} and \textit{GoldenGlobes} awards and the genres \textit{Drama}, \textit{Comedy} and \textit{Romance} look very similar, further suggesting a strong link between these awards. We then brush the +1 and +2 column in the \textit{Academy} award marginal histogram as shown in Figure \ref{fig:movies_analysis_patterns_brushed} (suppl. material) to confirm this finding.

The \textit{Berlin} award is mostly given to \textit{drama} or \textit{drama} and \textit{romance}. No other award is so restrictive in terms of genres.
In the heatmap cells for the \textit{SF} award we see that the most common combination of genres when winning only this award are \textit{Sci-Fi}, \textit{Action} and \textit{Adventure}.
We then brush the cell  \textit{SF} \textit{Action} +2 as shown in Figure \ref{fig:movies_analysis_patterns_brushed_sf} (suppl. material) to confirm this finding.
\par 
We can further refine the bivariate analysis drill-down into the \textit{Academy} / \textit{Drama} co-occurrences by selecting this cell (blue rectangle in Fig.~\ref{fig:movies_analysis_patterns}).
The detailed histograms and heatmap (Figure \ref{fig:movies_analysis_detail}) show co-occurrence distributions between all possible combinations.
In this case we are interested in the co-occurrence pattern with the highest cardinality, i.e., most awards and most genres.
The films that won an \textit{Acedemy} award with the highest genre cardinality can be found in the \textit{Drama} + 4 column of Figure \ref{fig:movies_analysis_detail}.
We can also see in the heatmap that this film won a \textit{SF} and \textit{BAFTA} award.
When hovering over the cells in this column we see that only one film is present here ("The Empire Strikes Back").
The films that won an \textit{Acedemy} award that also won the most other awards can be found in the \textit{Acedemy} + 4 row of Figure \ref{fig:movies_analysis_detail}.
We can identify two films here, one that also won \textit{Golden Globes}, \textit{MTVMovie}, \textit{BAFTA} and \textit{SF} awards with a genre mix of \textit{Drama} and \textit{Crime} ("Pulp Fiction"), and 
another, that also won \textit{GoldenGlobes}, \textit{Berlin}, \textit{BAFTA} and \textit{SF} awards with a genre mix of \textit{Drama} and \textit{Thriller} ("The Silence of the Lambs").

\section{Scalability of CrossSet}
When dealing with set-typed data, scalability is an important issue.
It is related to both an increase in data items, and to an increase in set elements~-- in a way similar to the scalability of $n$D visualization techniques with respect to the number of rows vs.\ the number of columns. 
\par 
For a frequency-based visualization, more data items rarely pose a technical challenge. 
Aggregation can be time-consuming, but clever algorithms and parallelization can help for really large counts. 
The visualization itself usually scales well, once everything is computed. 
A large number of set elements is a scalability challenge for the overall design. 
The number of possible combinations of set elements increases rapidly with cardinality, even in case of a single set-typed dimension.
\par 
CrossSet scales well to increasingly many data items. 
The size of the CrossSet view grows with the cardinality of the set-typed dimensions. 
It is not only that the number of cells in the heatmap grows, but each heatmap has more rows and columns to represent additional elements. 
Shrinking a heatmap cell to less than a pixel deteriorates the visualization. Practically, we should limit the minimum heatmap cell size to $4\times4$ pixels, or so, to avoid missing relevant patterns. Considering common screen sizes, this means that CrossSet scales to about 100 set elements in the best case. 
Up to 50 elements allow for an efficient work on a common high resolution screen. 
At this upper limit, however, we only use the top-most overview level of CrossSet. 
\par 
To further improve the scalability of CrossSet, we allow to merge rows and columns with a cardinality larger than a certain threshold. 
Furthermore, it is also possible to collapse certain elements to a single row or column by simple clicking on their titles (Figure~\ref{fig:movies_overview} (left)). Such a strongly aggregated view can serve as a starting point into a more detailed analysis. 
Some of the single set-typed dimension visualization techniques offer the user to select which elements should be analyzed. 
All others are then ignored. 
This can be considered as a scalabilty solution. 
So far, we've not realized such an approach in CrossSet, but there is no principle obstacle to doing so.  
Importantly, CrossSet is used to analyze two set-typed dimensions and their interaction, facing additional scalability challenges when compared to 1D approaches.  
\par 
During one of many analysis sessions we have analyzed near-Earth objects (NEOs), including comets and asteroids which move on orbits that brings them near Earth. 
They are of great interest to research as they can impose a major hazard if any such non-small NEO would impact Earth. 
NASA and the JPL Center for Near-Earth Object Studies~\cite{cneos} provide detailed information on NEOs. 
Their NEO data contains typical orbital dynamics information, including perihelion, aphelion, eccentricity, etc., as well as approach data (approach date, relative velocity, etc.), more data about the NEO itself (estimated size, absolute magnitude, etc.), and whether they are considered a hazard. 
Studying this dataset (4687 NEOs, 35 dims.) is challenging, not at the least due to the heterogeneous nature of all included information.
We derived two sets of binary indicators: for 20 data attributes, we asked whether the NEO's value is special, for ex., particularly large or small, etc. (set \#1), and for the same 20 attributes, we asked whether the NEO's value is common or uncommon (set \#2).
With CrossSet, we could then oppose these two types of indicators for all 20 dimensions in one overview (see Figure~\ref{fig:NEOsSuppl} in the suppl.\ material). 

\par 
Table~\ref{table:scalability_1} compares visualization techniques for set-typed data and their scalability. 
GridSet is an outlier regarding the number of set elements (196) and data items (2.5k).
It focuses on the depiction of individual elements, so it does not scale particularly well with many data items. 
It excels, on the other hand, on large sets. 
All other techniques utilize an aggregation approach which scales well with the number of data items.
They all do not scale even close to as GridSet does. 
Still, they all proved their usefulness in a wide variety of applications and domains, possibly due to the fact that in many cases, even when the number of elements is large, the individual data items have relatively small subsets of them.  
CrossSet scales similarly as other aggregation-based techniques. 
Its main goal was not to extend scalability, but to make it possible to cope with two set-typed dimensions simultaneously and to enable a bivariate analysis. 
\begin{table}
\setlength{\tabcolsep}{0.75pt}
\caption{A comparison of set visualization techniques, based on the number of data points and the number of sets used and scalability limitations (stated by the authors or derived from the used examples).}
\footnotesize{\phantom{Xxx}\vspace{-7ex} 
\begin{center} 
\begin{tabular}{@{~}l@{~}l@{~}r@{~~}c@{~}c@{~~~}c@{~}c@{}c@{}c@{}} 
\multirow{2}{*}{\textbf{visualization}}
& \multirow{2}{*}{\textbf{dataset\ analyzed}}
& \textbf{\# data}
& \multicolumn{2}{l}{\textbf{\# set-typed}} 
& \multicolumn{2}{l}{\textbf{univar.}} 
& \multicolumn{2}{l}{\textbf{bivar.}} 
\\[-1ex] 
& 
& \textbf{points}
& \textbf{~dim.}
& \textbf{elem.}
& \textbf{~1~} 
& \textbf{2 sets~}
& \textbf{~1~~~}
& \textbf{2+~~~~} 
\\
\midrule
Set'o'gram~\cite{SetOGram2008} 
& ~CRM data 
& 93k 
& 1 
& 10 
& \checkmark
& \checkmark
& 
& \\
\hline
\multirow{2}{*}{RadialSet \cite{Alsallakh-2013-a}}
& ACM classification 
& 50k
& 1
& 11
& \multirow{2}{*}{\checkmark}
& 
& 
& \\
& IMDb movies 
& 525k 
& 2
& 28, 35
& 
& 
& 
& \\
\hline
\multirow{3}{*}{UpSet~\cite{UpSet_2014}}
& Macroeconomics
& 1354
& 1
& 8
& \multirow{3}{*}{\checkmark}
& 
& 
& \\[-0.5ex]  
& Genomic variation
& 15
& 1
& 4
& 
& 
& 
& \\[-0.5ex] 
& IMDb movies 
& 3883 
& 1
& 11
& 
& 
& 
& \\
\hline
OnSet~\cite{OnSet_2014} 
& Sharks blood
& 50 
& 1
& 7
& \checkmark
& 
& 
& \\
\hline
\multirow{4}{*}{PowerSet~\cite{PowerSet_2017}}
& Vehicles
& 200
& 1
& 4
& \multirow{4}{*}{\checkmark}
& 
& 
& \\[-0.5ex]  
& Laptops
& 836
& 1
& 12
& 
& 
& 
& \\[-0.5ex]  
& Survey
& 618
& 1
& 7
& 
& 
& 
& \\[-0.5ex]  
& IMDb movies
& 500k+ 
& 2
& 31, 189
& 
& 
& 
& \\
\hline
\multirow{5}{*}{AggreSet~\cite{Yalcin-2016-a}}
& Courses
& 4300 
& 1
& 83
& \multirow{5}{*}{\checkmark}
& 
& 
& \\[-0.5ex]  
& Ingredients
& 5k
& 1
& 82
& 
& 
& 
& \\[-0.5ex]  
& Record types
& 284
& 1
& 18
&
& 
& 
& \\[-0.5ex]  
& Le Miserables
& 365
& 2
& 6, 80
&
& 
& 
& \\[-0.5ex]  
& IMDb movies
& 3884 
& 1
& 17
& 
& 
& 
& \\
\hline
\multirow{2}{*}{GridSet~\cite{GridSet_2020}}
& Militarized dispute
& 2586
& 1
& 196
& \multirow{2}{*}{\checkmark}
& 
& 
& \\[-0.5ex]  
& IMDb movies 
& 5k 
& 1
& 26 
& 
& 
& 
& \\
\hline
DualRadial 
& \multirow{2}{*}{IMDb movies}
& \multirow{2}{*}{1590}
& \multirow{2}{*}{2}
& \multirow{2}{*}{7, 15}
& \multirow{2}{*}{\checkmark}
& \multirow{2}{*}{\checkmark}
& \multirow{2}{*}{\checkmark} 
& \\[-0.5ex]  
Set~\cite{Matkovic-2020-c}
& 
& 
& 
& 
& 
& 
& 
& \\
\hline
\multirow{4}{*}{CrossSet}
& Drives 
& 250k+ 
& 2
& 5, 5
& \multirow{4}{*}{\checkmark}
& \multirow{4}{*}{\checkmark}
& \multirow{4}{*}{\checkmark}
& \multirow{4}{*}{\checkmark} \\[-0.5ex]  
& Patients 
& 5434 
& 2
& 5, 5
& 
& 
& 
& \\[-0.5ex]  
& IMDb movies
& 1521 
& 2
& 6, 10
& 
& 
& 
& \\[-0.5ex]  
& Near-Earth objects
& 4687 
& 2
& 20, 20
\end{tabular}
\end{center}
\label{table:scalability_1}
}
\end{table}
\section{Discussion}
The analysis of two set-typed dimensions is challenging and imposes a significant cognitive load. 
To still manage, we decomposed CrossSet into components, so that the user can drill down from the overview into details step-by-step. 
\par 
We explored a large number of different design alternatives, including such where all information would be encoded in the top level bins, but the resulting design did not scale.
In a cross-section of two elements, we have to show the number of items for each cardinality combination. 
Encoding the combinations themselves would not scale at all.
In a usual case of multivariate data where each data item has additional attributes, the use of coordinated multiple views makes it possible to observe CrossSet patterns in relation to other attributes, as well. 
The here described design still leaves room for further development.
For some users an organization going from top\,/\,left towards bottom\,/\,right could be more intuitive. 
Assigning different color scales to different levels of detail could also facilitate certain special cases.
Still, the here presented design was appropriate to enable a successful bivariate analysis of set-typed attributes that was not possible before. 
We acknowledge that the learning curve for CrossSet is steep. However, the analysis of bivariate set-typed data is inherently complex and multi-layered, making it practically impossible to devise a simple solution.
Promising directions of future research include providing user guidance (which will also lower the barrier to entry)  and comparison of different modes (item- or element-based, value- or rank-based, etc.).  
A more formal evaluation with domain experts in several fields is also planned. 

\section{Conclusions}
In this paper, we describe CrossSet, a first step towards a true bivariate analysis of data with two set-typed dimensions. 
CrossSet supports an analysis at different levels of aggregation and through different perspectives: attribute properties vs.\ data items, value-based vs.\ rank-based, actual counts vs.\ their deviation from an even ``default'' distribution, etc.  
CrossSet relies heavily on interaction to handle the different levels of detail required when analyzing such data.
\par 
We carefully identified the most important general requirements for analyzing two set-typed dimensions.
Such data scenarios are common, but their thorough bi-\,/\,multivariate analysis is highly challenging. 
But in many cases, it would even be beneficial to study more than two set-type dimensions, e.g., for medical data containing sets for symptoms, diseases and treatments.
Revealing correlations between three set-typed dimensions requires a substantial amount of further research. 
\par 
Data attributes can also be converted into a set-typed form, amounting to an interesting strategy for dimension reduction. 
Several Boolean attributes are a clear candidate, as demonstrated in the NEOs case.
Supporting multiple set-typed dimensions makes it possible to group related attributes. 
In case of multiple numerical dimensions, one can use thresholds, and then convert these dimensions to Boolean attributes, followed by a grouping into set-typed dimensions. 
We plan to explore this idea more thoroughly in future work.  
\par 
The research on multiple set-typed dimensions is in its pioneering stage.
While CrossSet does not master all related challenges, it appears to be a first and important step towards a comprehensible understanding of the analysis requirements for complex data that contains multiple set-typed dimensions. 
According to our experience, sets are a natural choice for representing relevant information and we do need analysis methods especially tailored for such data in order to fully exploit related analysis potentials.

\section*{Supplemental Materials}
\label{sec:supplemental_materials}
The full paper, including all supplemental materials, is available on arXiv at \url{https://xxxxxxxx}.
The appendix includes additional figures in Sect.~\ref{sec:AddFigs} (Figs.~\ref{fig:probdemoCrossSet02}--\ref{fig:NEOsSuppl}), including enlarged versions of figures from the main paper, as well as additional figures providing additional visualizations. All supplemental figures are referred to in the main text. In Sect.~\ref{suppl:additinalCase}, with three additional figures (Figs.~\ref{fig:covid_four}--\ref{fig:covid_compare}), we present an additional demonstration case, complementing Sect.~\ref{sec:evaluation}. 
%
%
%
%
%
\section*{Acknowledgments}
We thank Hemanth Hari, who collected and provided the Covid19 data. 
We also thank Michael Beham and Elena Ginina for numerous discussions and an early implementation. 

The VRVis GmbH is funded by BMK, BMAW, Tyrol, Vorarlberg and Vienna Business Agency in the scope of COMET - Competence Centers for Excellent Technologies (911654) which is managed by FFG.

Parts of this work have been supported by the Virginia Tech Institute for Creativity, Arts, and Technology.

Parts of this work have been done in the context of CEDAS, i.e., 
\mbox{the Center for Data Science at the University of Bergen, Norway}.%

\bibliographystyle{abbrv-doi-hyperref}
\bibliography{CrossSet_IEEEVIS_2025}           
\appendix\newpage%
\onecolumn\large 
\begin{center}\Large\paptit\end{center}
\begin{center}\huge{}Supplementary Material\end{center}
\vspace{2ex}\par\bigskip\noindent 
As supplementary material, we provide additional figures (Sect.~\ref{sec:AddFigs})~-- also referred to from the paper~-- as well as another demonstration case (Sect.~\ref{suppl:additinalCase}) to further substantiate our presentation of CrossSet.
\vspace{2ex}\section{Additional Figures}\label{sec:AddFigs}
\begin{figure}[h!]
    \centering
    \includegraphics[width=\linewidth]{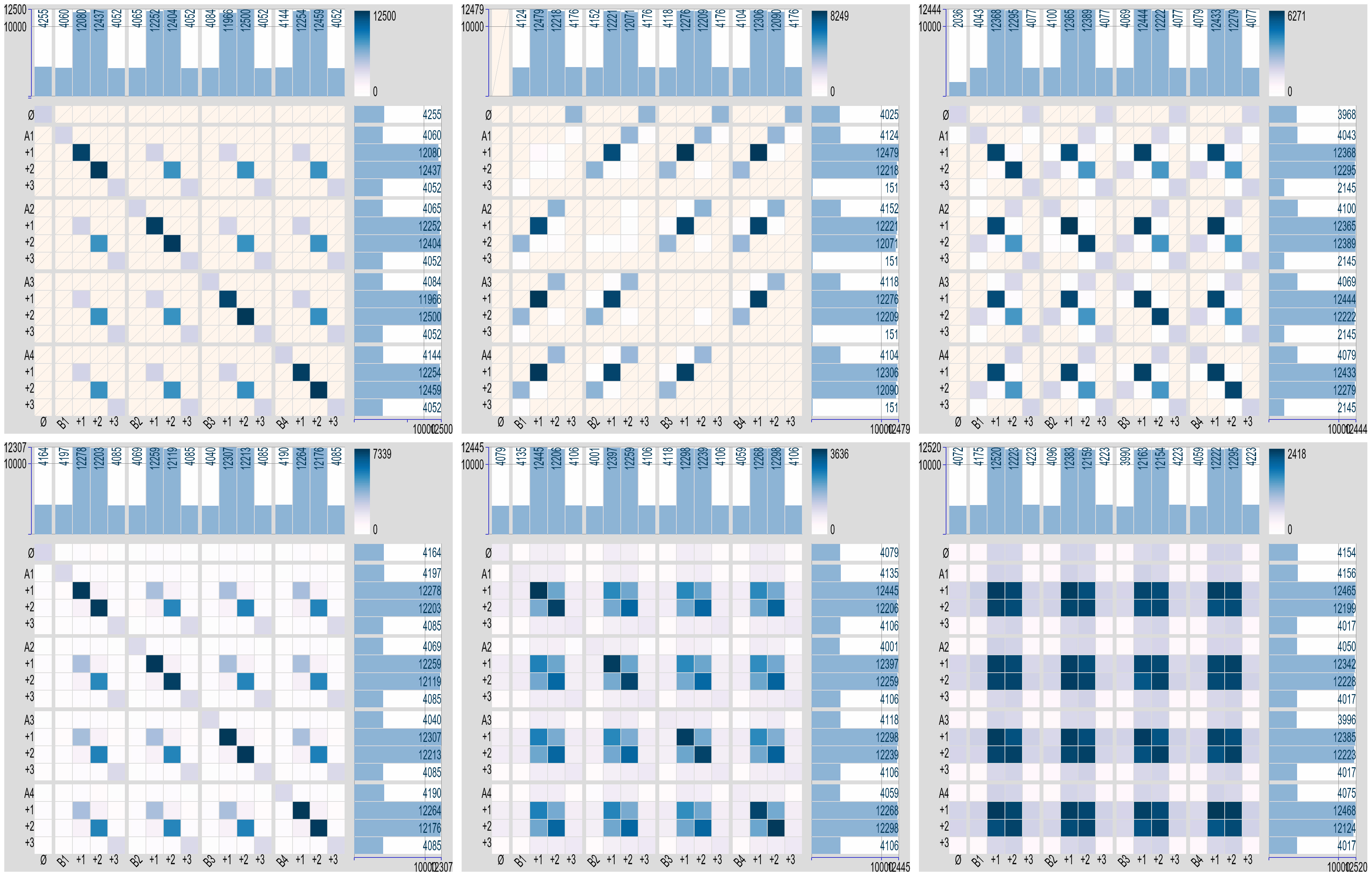}
    \caption{CrossSet showing different bivariate association structures in the synthetic datasets \textbf{S1}, \textbf{S2}, \textbf{S3}, \textbf{S4}, \textbf{S5}, and \textbf{S6}, which cannot be distinguished by considering the two set-typed dimensions individually (see Fig.~\ref{fig:histAB}).}
    \label{fig:probdemoCrossSet02} 
\end{figure}
\begin{figure}[t]
\begin{center}
\includegraphics[width=\linewidth]{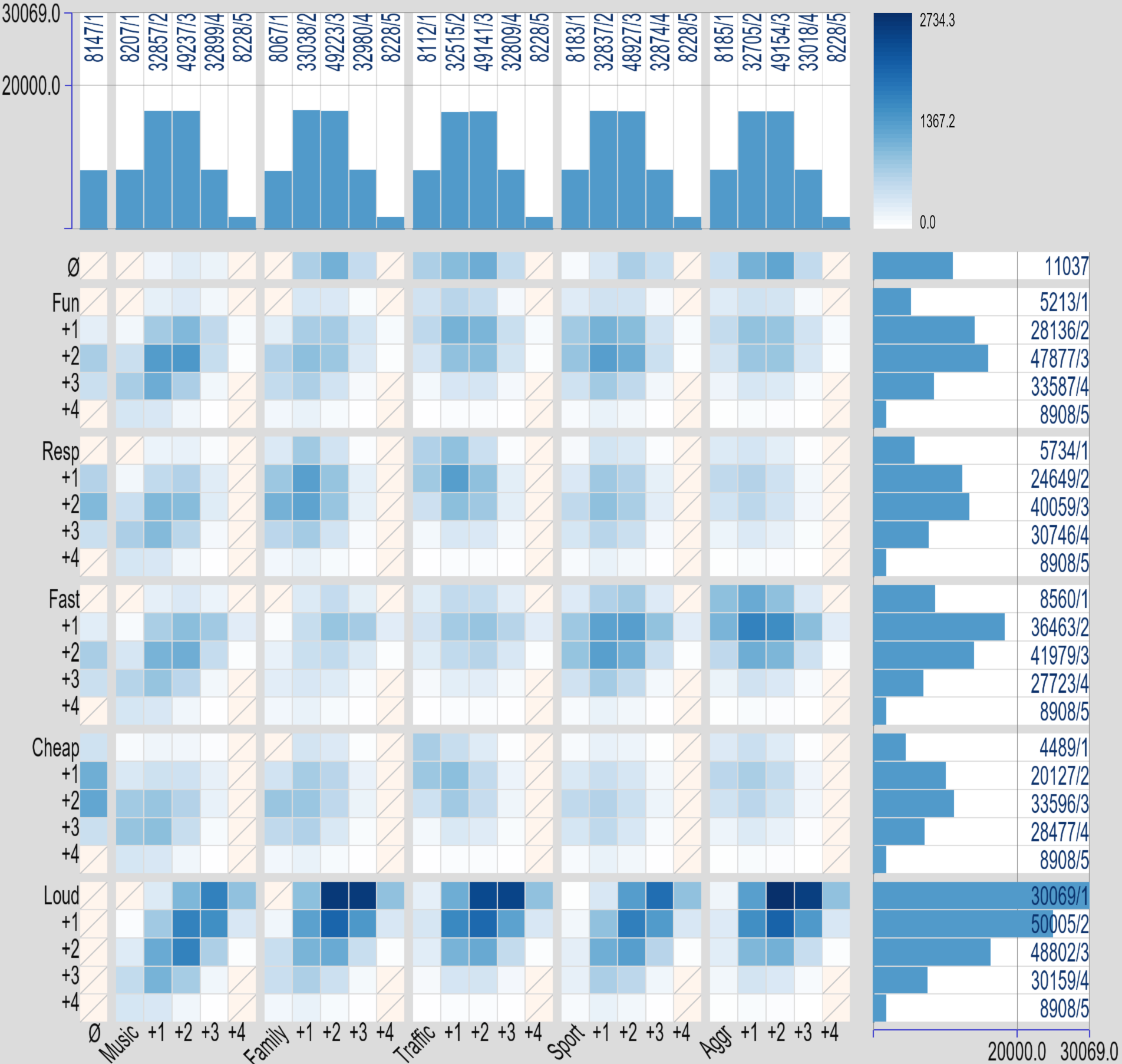}
\caption{The basic visual encoding of the drives dataset. The \textit{Loud} element comes most often (see marginal histograms). It also comes alone more often than in other combinations. The co-occurrence pattern of \textit{Fast} vs.\ \textit{Sport} appears similar to that one of \textit{Fast} vs.\ \textit{Fun}.}
\label{fig:drive_all_1}
\end{center}
\end{figure}
\begin{figure}[t]
\begin{center}
\includegraphics[width=\linewidth]{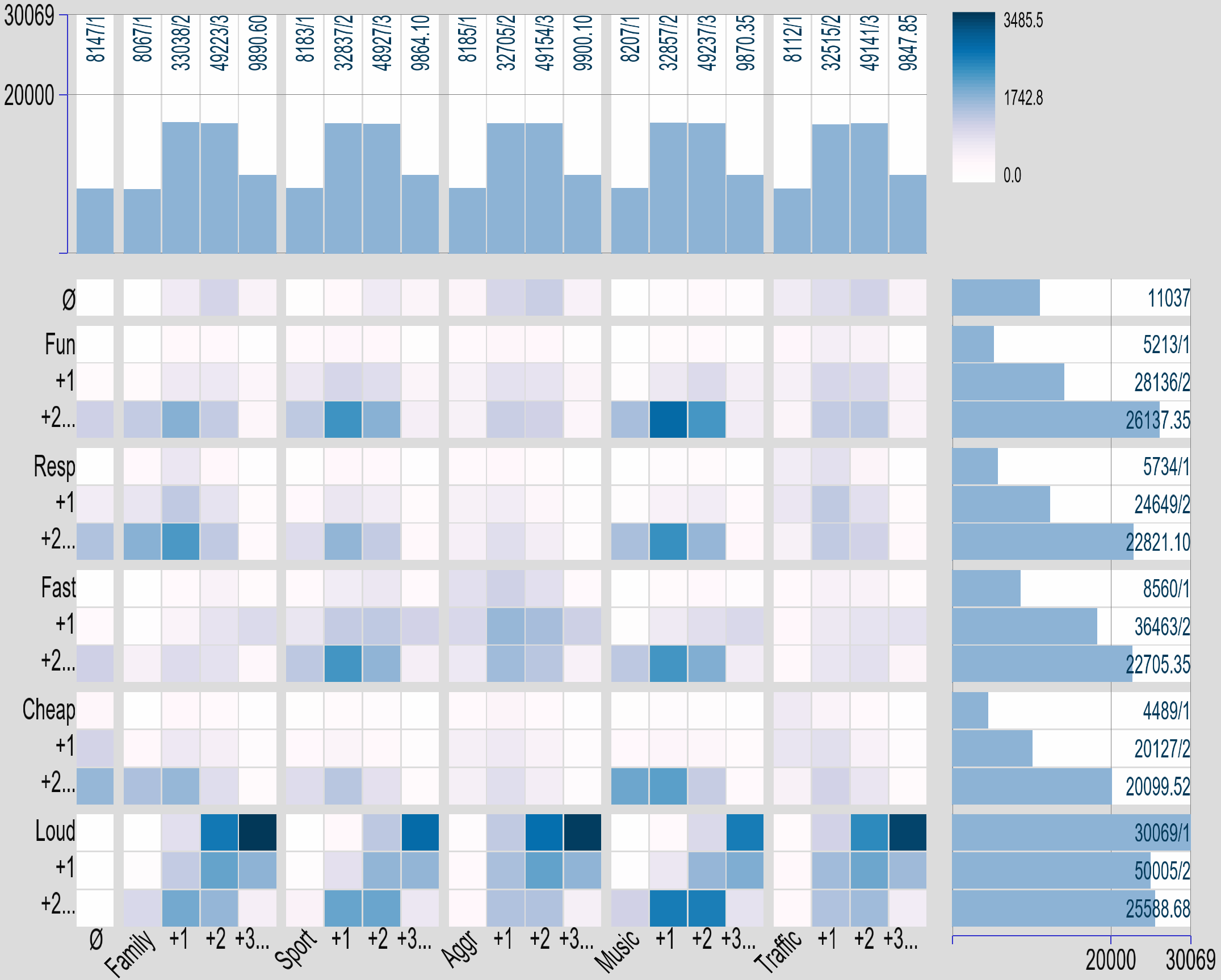}
\caption{CrossSet showing the drives dataset with limited cardinalities. The \textit{Input} set elements are shown up to cardinality four. Note that the "+3.." columns also include the subsets with cardinality five. The \textit{Output} set elements are shown up to cardinality three. The "+2.." rows also include the subsets with cardinality four and five. Also note that the marginal histograms show decimals instead of fractions for the rows / columns that show combined cardinalities.
}
\label{fig:partiallyCollapsed}
\end{center}
\end{figure}
\begin{figure}[t]
\begin{center}
\includegraphics[width=\linewidth]{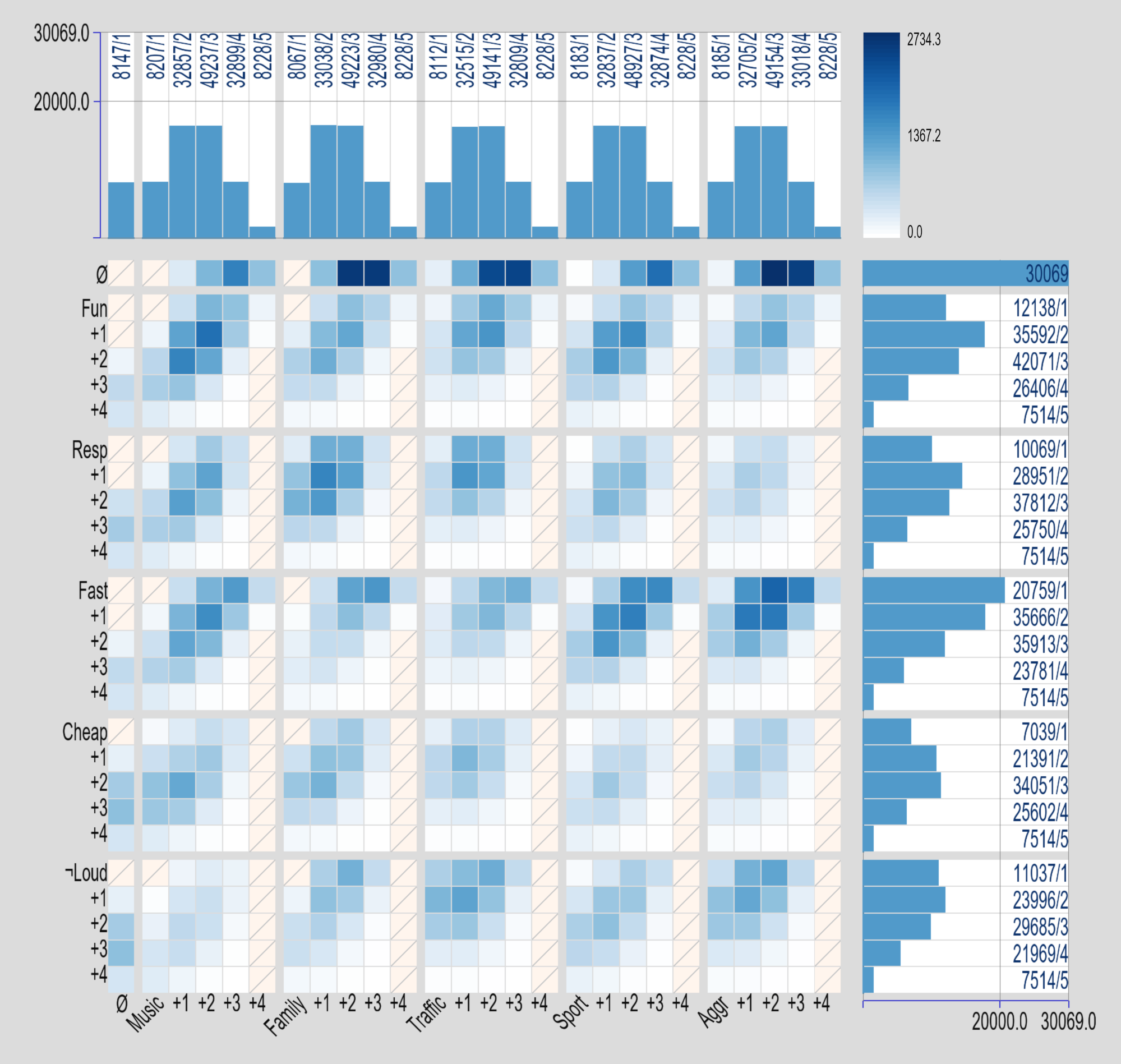}
\caption{CrossSet in the same configuration as in Figure~\ref{fig:drive_all_1} but with  {\textit{Loud}} inverted, indicated by label $\neg$\textit{Loud}, which can be interpreted as \textit{Silent}.  $\neg$\textit{Loud} is added to all sets that did not have \textit{Loud}, and  \textit{Loud} is removed from all sets that had it. Note the increased prevalence of empty sets  (all sets that had \textit{Loud} alone turned into $\emptyset$,  \textit{Fast} appears most prominent, instead.}
\label{fig:silent}
\end{center}
\end{figure}
\begin{figure}[t]
\begin{center}
\includegraphics[height=.9\textheight]{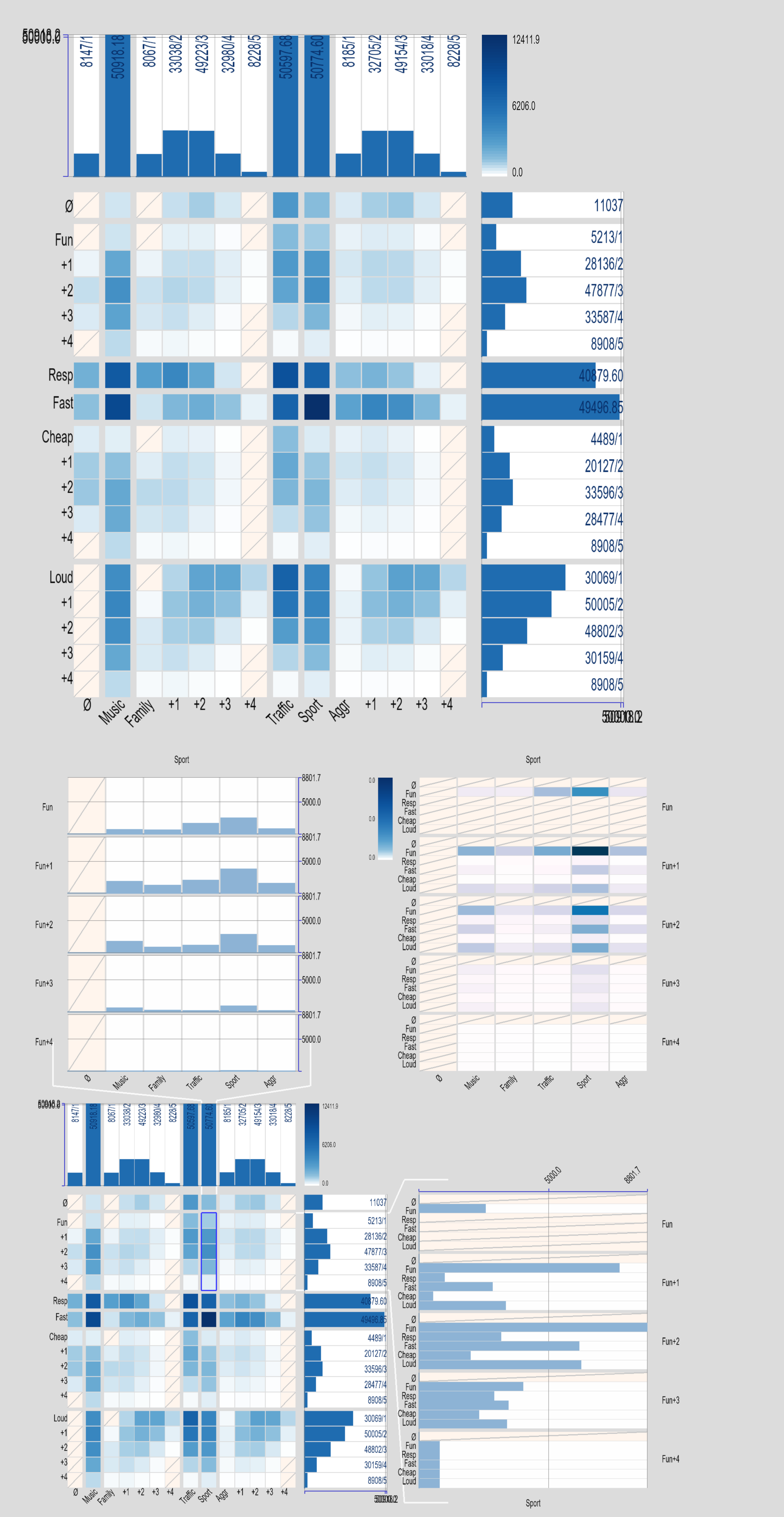}
\caption{\textbf{Top}: The main heatmap of the CrossSet with two collapsed rows, \textit{Resp} and \textit{Fast}, and three collapsed columns, \textit{Music}, \textit{Traffic}, and \textit{Sport}. Collapsed heatmaps show the sum of all cells. Collapsing is used to reduce the complexity and size of the view, and to support focusing on details. \textbf{Bottom}: When a collapsed heatmap is selected, \textit{Fun} and \textit{Sport} here, the detail histograms on the top and right, and the detail heatmap on top right, are also collapsed to match the originally selected heatmap in the main view.}
\label{fig:collapsingExamples}
\end{center}
\end{figure}
\begin{figure}[t]
\begin{center}
\includegraphics[width=\linewidth]{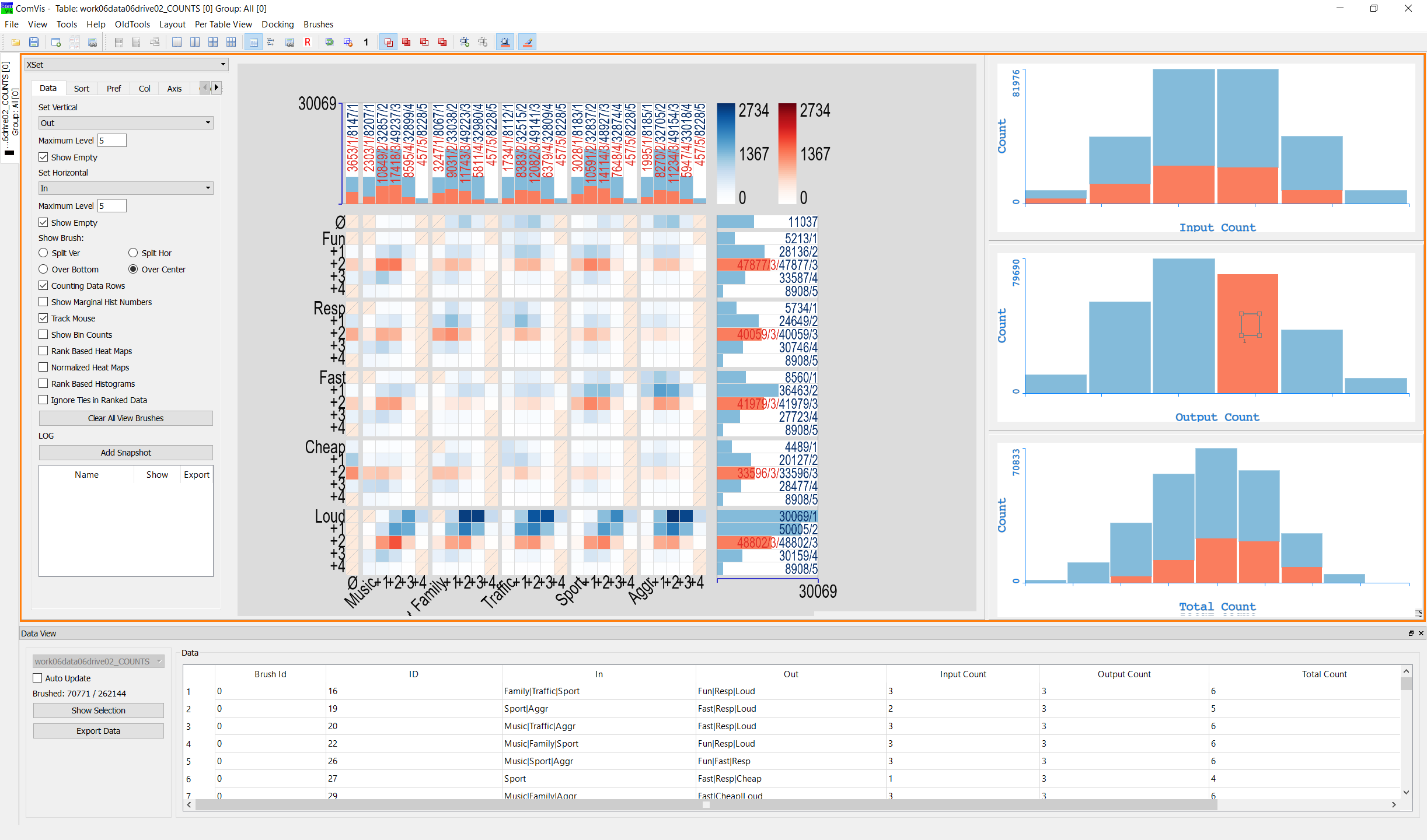}
\caption{ CrossSet is integrated into a coordinated multiple views (CMV) system. In this example, the user has brushed a bin in the histogram view (right, middle), which shows the cardinality of Output. Since all views are linked, CrossSet and the two other histograms in this setup also show the brushed data items accordingly. Additional views provide alternative perspectives on the data and allow the user to brush data in the context of these perspectives. The system can be configured to display different attributes using various views (e.g., scatter plots). Details can be shown on demand—for example, the currently brushed data items are displayed in the table at the bottom. The control pane on the left of the CrossSet view is used to configure and parametrize CrossSet. The user can change the counting modes, colors, order of elements, cardinality limits, and many more. Due to a large number of parameters, the control pane is divided into several tabs. }
\label{fig:CMV}
\end{center}
\end{figure}

\begin{figure}[t]
\begin{center}
\includegraphics[width=\linewidth]{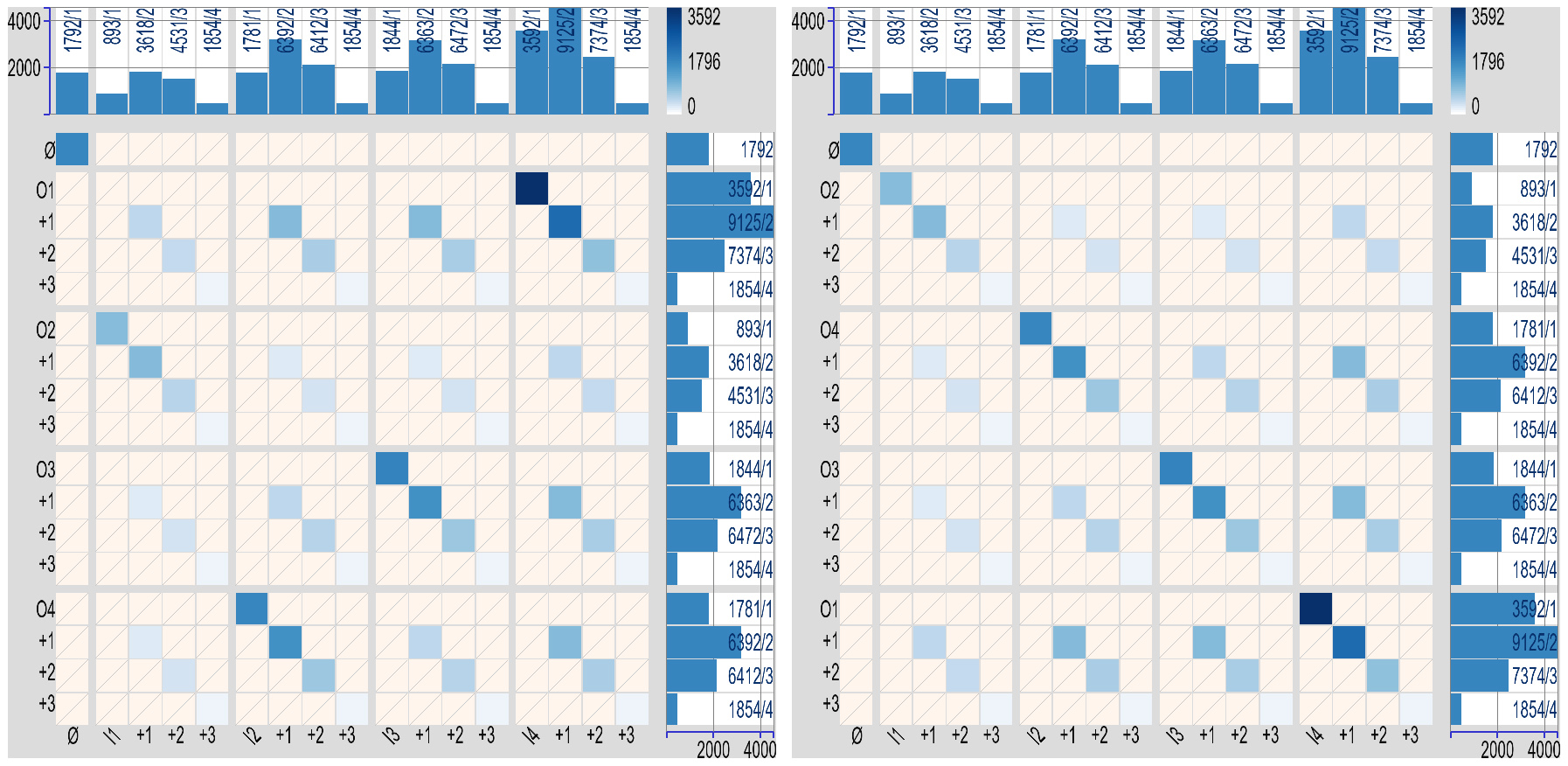}
\caption{A synthetic data set with a clear structure is shown. After reordering of the rows, the pattern is particularly clear.}
\label{fig:reorder_comparison}
\end{center}
\end{figure}
\begin{figure}[t]
\begin{center}
\includegraphics[width=\textwidth]{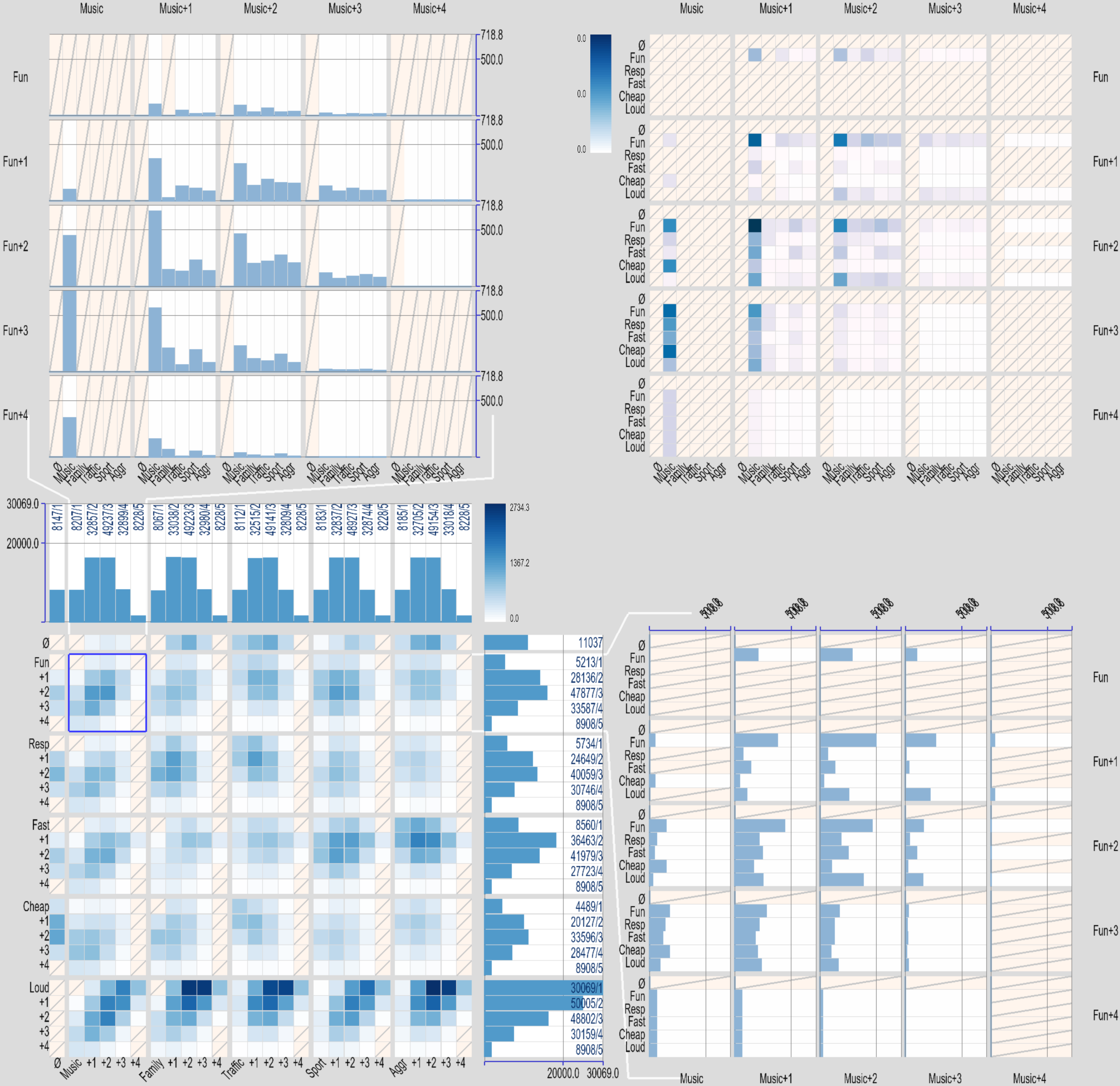}
\caption{A larger version of the Figure~\ref{fig:allDetails_First_small}. Detail histograms (top-left and bottom-right) and a detail heatmap (top-right) provide more detailed counts. For the selected heatmap in the lower-left (blue rectangle), one detail histogram is shown for each heatmap cell, providing insight into what the aggregations ``+1'', ``+2'', etc., stand for in the heatmap. 
The detail heatmap (top-right) relates these histograms to each other.}
\label{fig:allDetails_First_Large}
\end{center}
\end{figure}
\begin{figure}[t]
\begin{center}
\includegraphics[width=\linewidth]{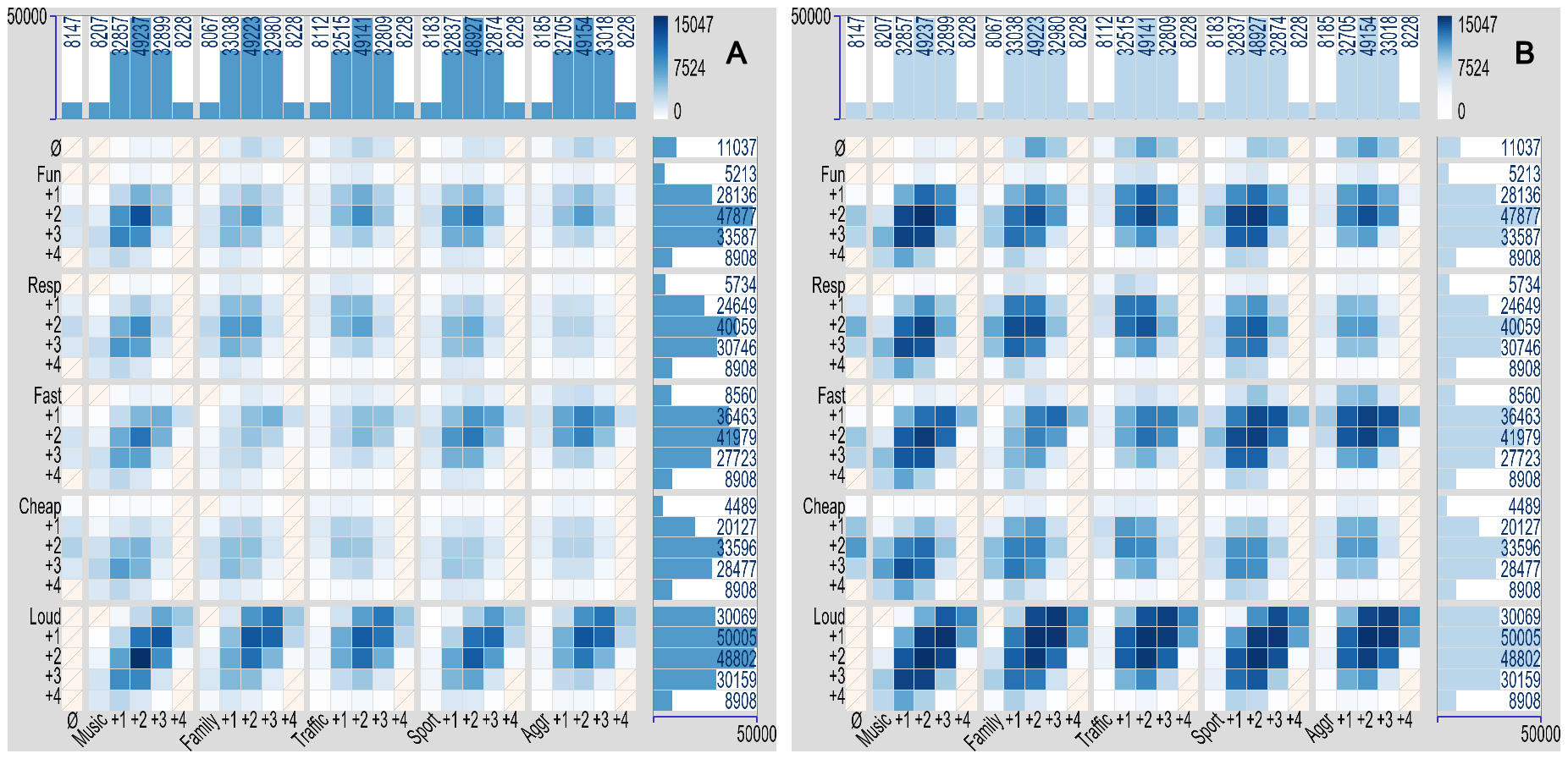}
\caption{Two additional counting modes in CrossSet.~ \textbf{Left}: The set elements are counted, not the data items. For example, if a data item (a drive, here) has a set of three elements for inputs and two for outputs, it will be counted six times.~ \textbf{Right}: In the rank-based mode, the bins are ranked according to the number of elements and assigned colors based on the rank, regardless of the element count.}
\label{fig:mappings}
\end{center}
\end{figure}
\begin{figure}[t]
\begin{center}
\includegraphics[width=\linewidth]{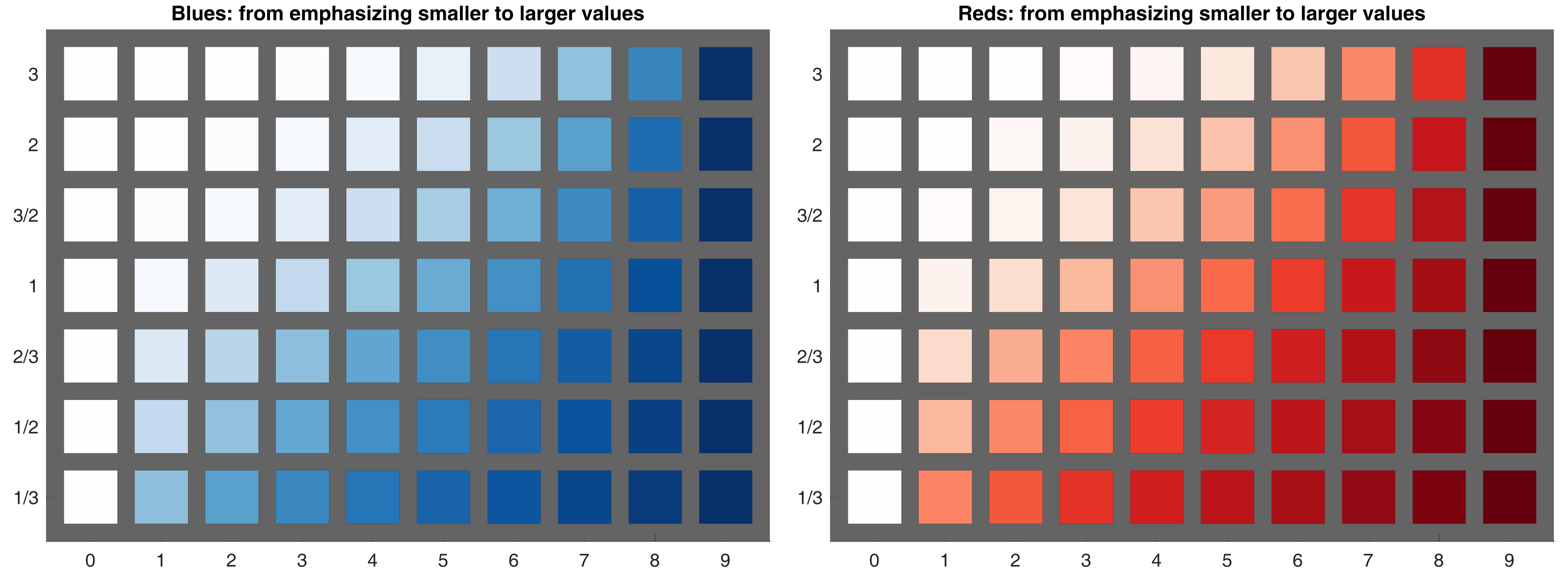}
\caption{Blue and red non-linear color scales as used in CrossSet. Given a skewed distribution of numbers, a non-linear color scale can help to achieve a better color contrast in the visualization. The color scales in the upper half stress differences among higher values, the middle row color scale is neutral (as linear as possible), the bottom color scales stress differences among lower values.}
\label{fig:colorScales}
\end{center}
\end{figure}
%
%
%
\begin{figure*}[t!] 
\centering
\includegraphics[width=\linewidth]{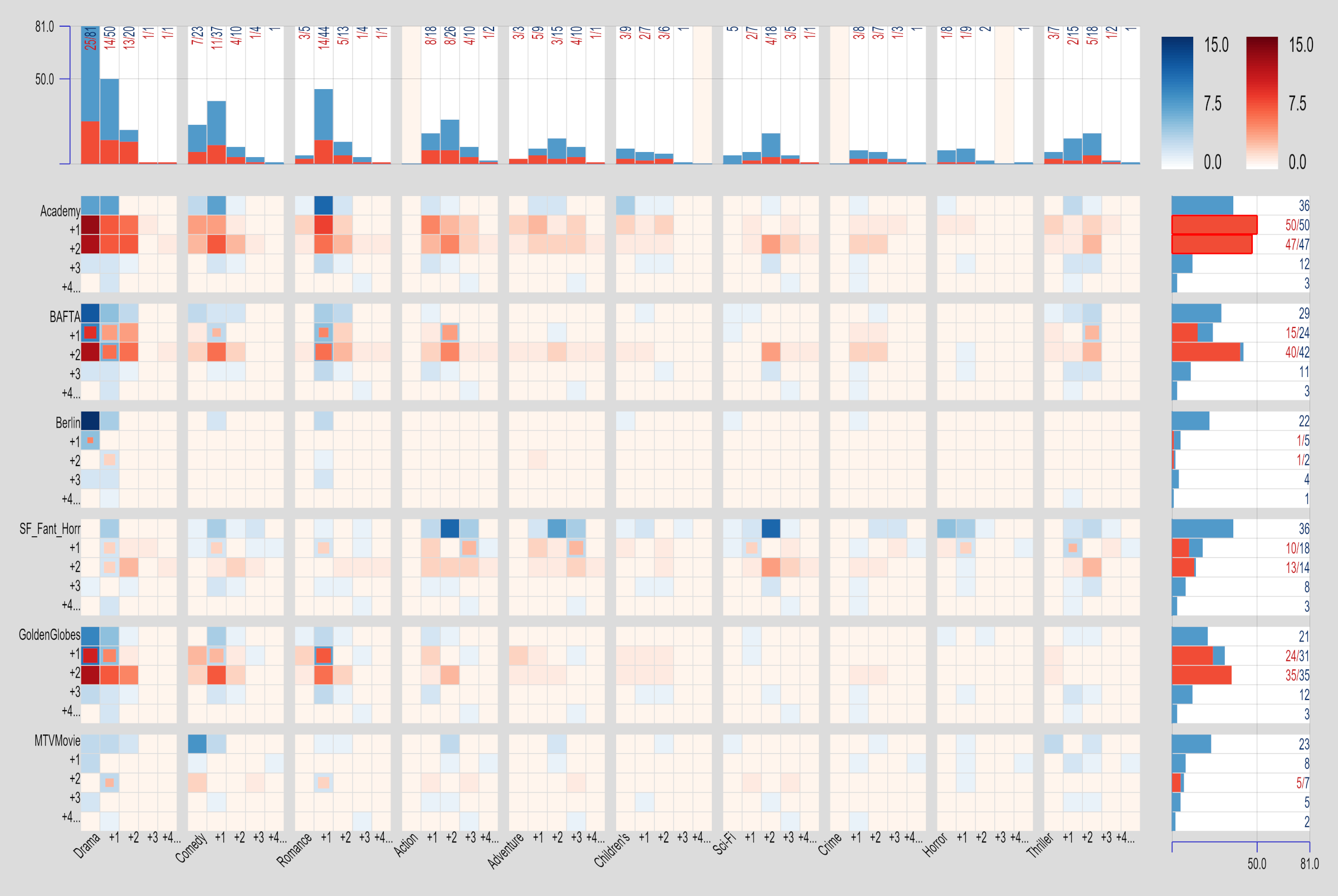}\\[-1.5ex] 
\caption{To confirm the strong link between \textit{Academy}, \textit{BAFTA}, and \textit{GoldenGlobes}, we brush the +1 and +2 columns in the \textit{Academy} award's marginal histogram. When looking at the +2 marginal histogram for awards we see that all \textit{GoldenGlobes} and nearly all \textit{BAFTA} winning films are highlighted.  
When looking at the +1 marginal histograms of the same awards we also see that the combination with either \textit{BAFTA} or \textit{GoldenGlobes} is very prominent.  
Also the \textit{SF} award has a strong link to the \textit{Academy} awards, but as we can see in the patterns of the highlighted cells in the heatmap the genre mix when winning \textit{Academy}, \textit{BAFTA}, and \textit{GoldenGlobes}, is very different to the case when the film won 
\textit{Academy}, \textit{BAFTA}, and \textit{SF}.}
\label{fig:movies_analysis_patterns_brushed}
\end{figure*}
\begin{figure*}[t!] 
\centering
\includegraphics[width=\linewidth]{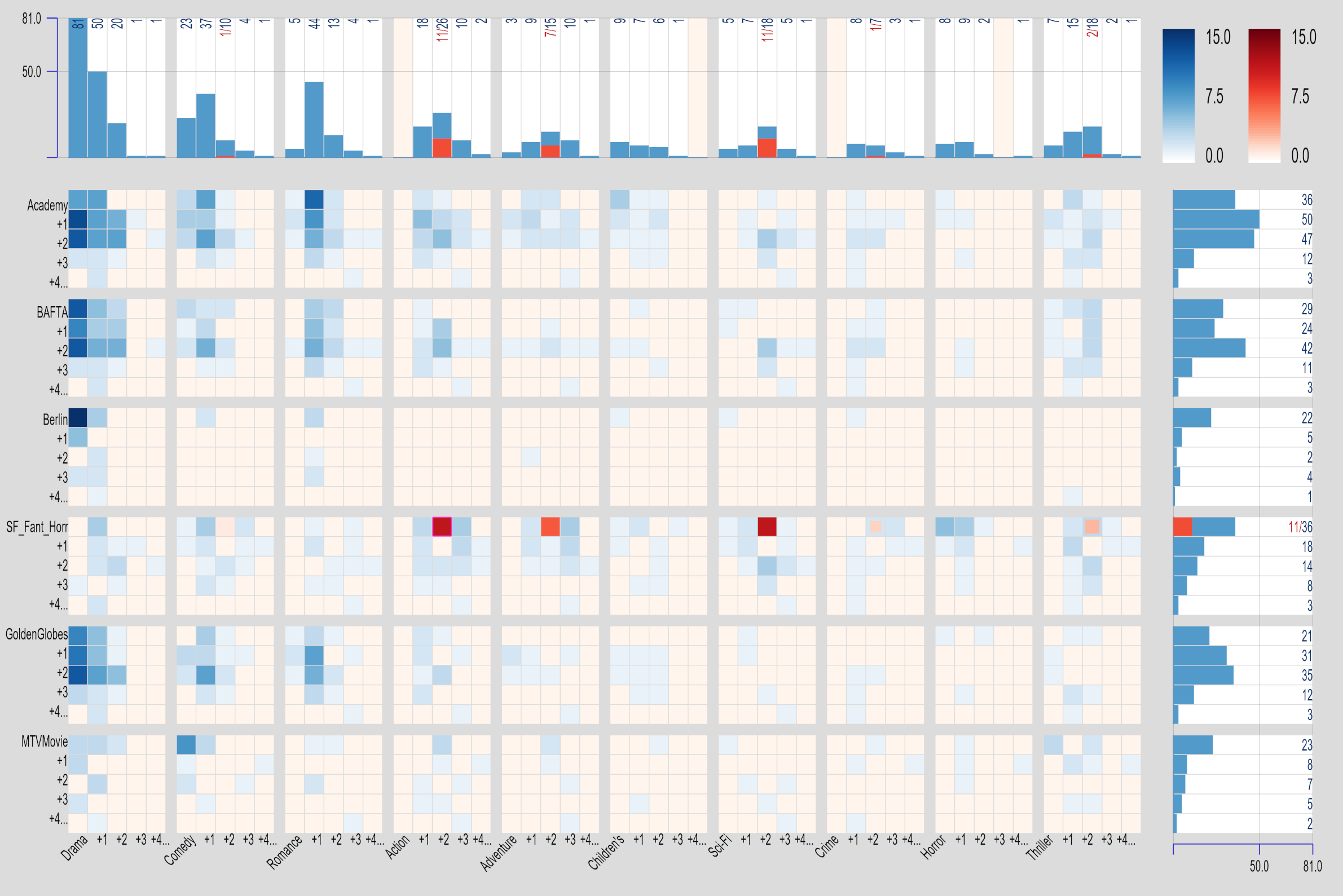}\\[-1.5ex] 
\caption{In the heatmap cells for the \textit{SF} award we see that the most common combination of genres when winning only this award are \textit{Sci-Fi}, \textit{Action}, and \textit{Adventure}.
Brushing the cell \textit{SF} vs.\ \textit{Action}+2, we can confirm this finding: the cells related to \textit{Sci-Fi} and \textit{Adventure} are fully highlighted.}
\label{fig:movies_analysis_patterns_brushed_sf}
\end{figure*}
\begin{figure*}[t!] 
\centering
\includegraphics[width=\linewidth]{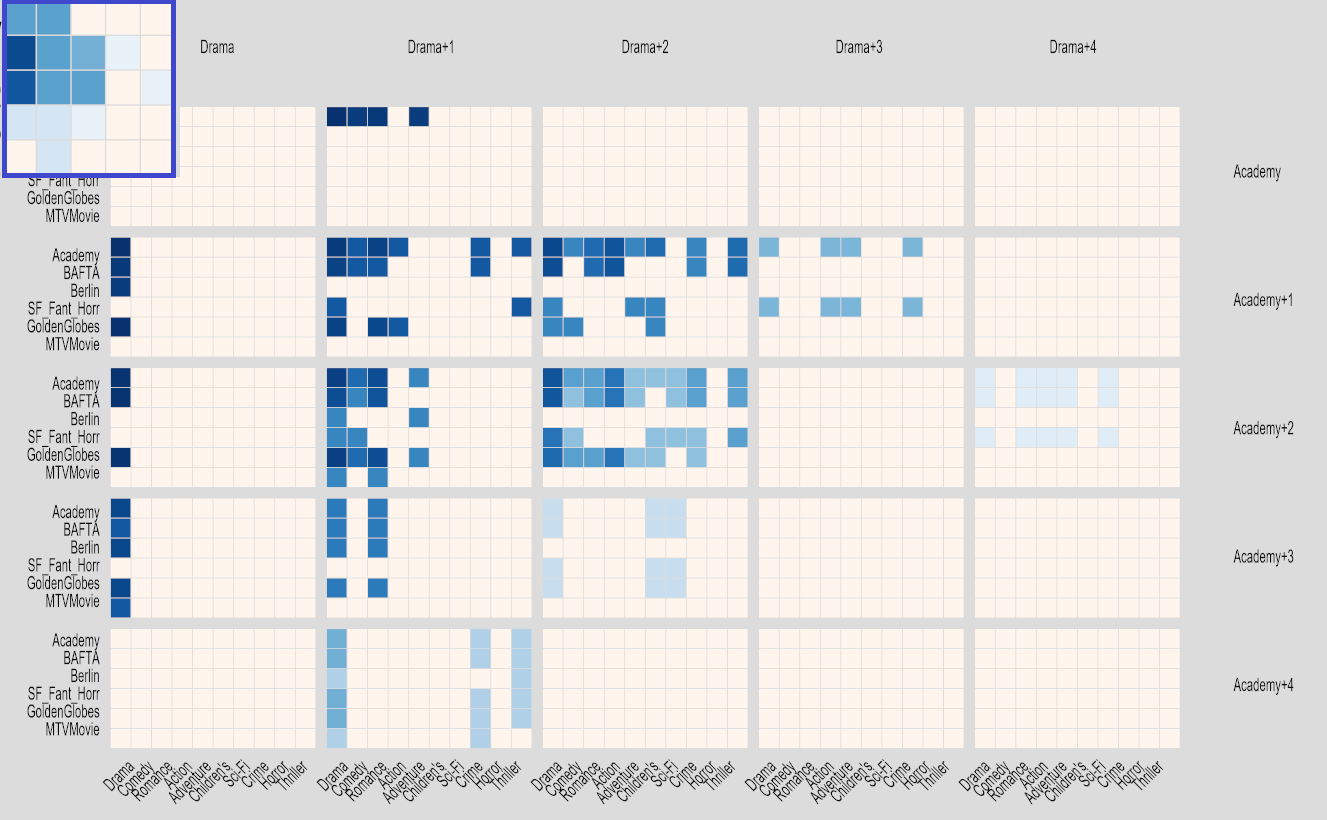}\\[-1.5ex] 
\caption{The detailed heatmap of the cell \textit{Academy} / \textit{Drama} shows the co-occurrence distributions between all possible combinations within the cell.
The cell of the main view that was selected to show this detail view is shown on the top-left.
In this case, we are interested in the co-occurrence pattern with the highest cardinality, i.e., most awards and most genres.
The films that won an \textit{Acedemy} award with the highest genre cardinality can be found in the \textit{Drama}+4 column.
We can also see in the heatmap that this film won a \textit{SF} and \textit{BAFTA} award (``The Empire Strikes Back'').
The films that won an \textit{Acedemy} award that also won the most other awards can be found in the \textit{Acedemy}+4 row.
We can identify two films here, one that also won \textit{Golden Globes}, \textit{MTVMovie}, \textit{BAFTA}, and \textit{SF} awards, with a genre mix of \textit{Drama} and \textit{Crime} (``Pulp Fiction''), and another one, that also won \textit{GoldenGlobes}, \textit{Berlin}, \textit{BAFTA}, and \textit{SF} awards, with a genre mix of \textit{Drama} and \textit{Thriller} (``The Silence of the Lambs'').}
\label{fig:movies_analysis_detail}
\end{figure*}
%
%
%
\begin{figure*}[t!] 
\centering
\includegraphics[width=\linewidth]{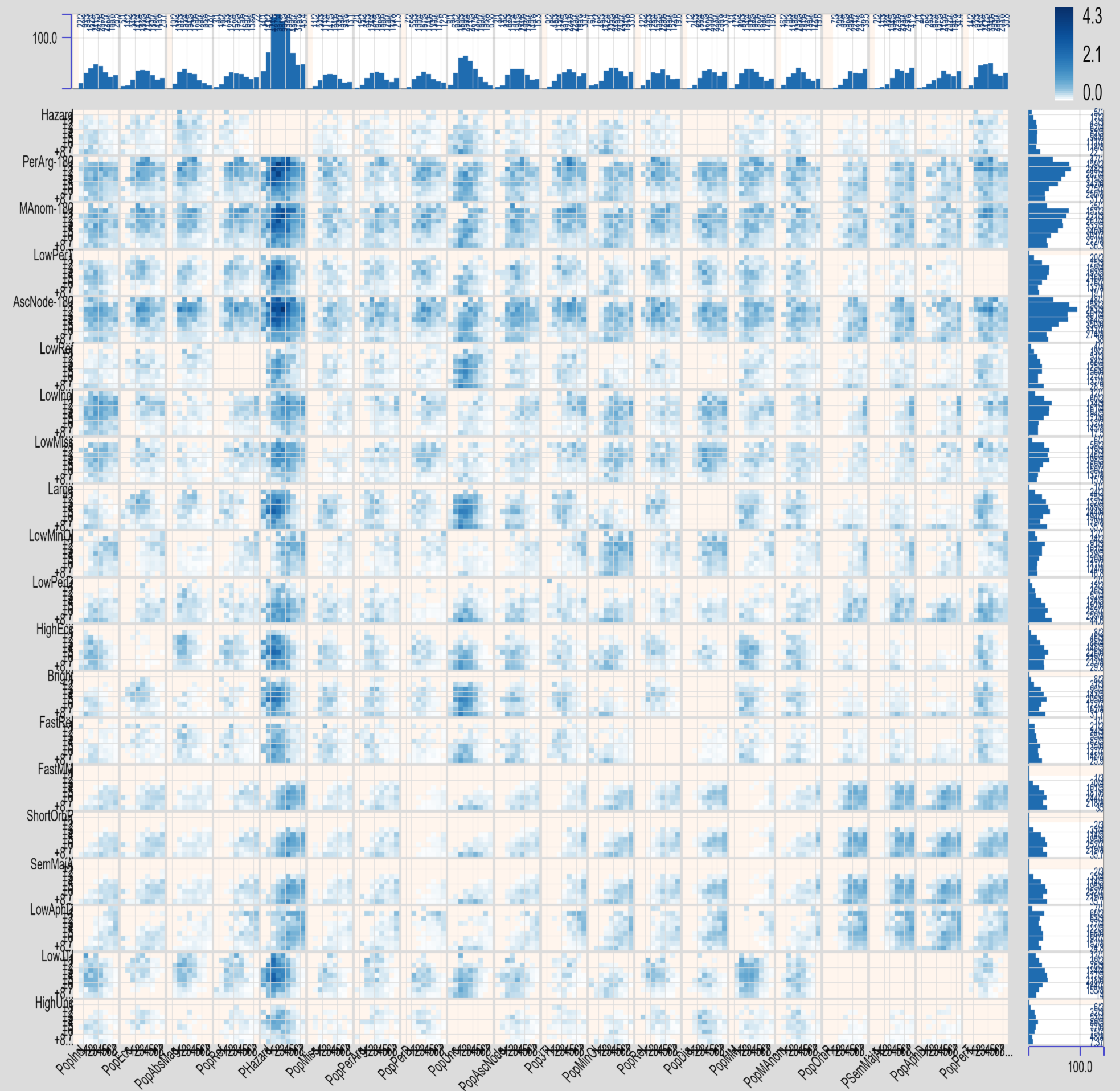}\\[-1.5ex] 
\caption{CrossSet showing NASA data about $4\,687$ Near-Earth Objects (NEOs) in $20\!\times\!20$ heatmaps, relating 20 value-based indicators, categorizing, for example, orbital properties, to 20 prevalence-based indicators. Showing 400 heatmaps with 81 cells each demonstrates the scalability of CrossSet: while the mouse-over function of CrossSet, combined with interactive highlighting, helps when it comes to details, we still get a good overall overview of the entire dataset.}
\label{fig:NEOsSuppl}
\end{figure*}
\clearpage
%
%
\section{Additional Demonstration}
\label{suppl:additinalCase}
\setlength{\footnotesep}{4mm}
This second demonstration case refers to a patients dataset\footnote{~ H. Hari:~ Symptoms and Covid presence,~ \texttt{https://www.kaggle.com/hemanthhari/symptoms-and-covid-presence/}} that contains 5\,434 patient records with information about the patients' symptoms, diseases, and their Covid19 status~.
\begin{figure*} 
\centering
\includegraphics[width=\linewidth]{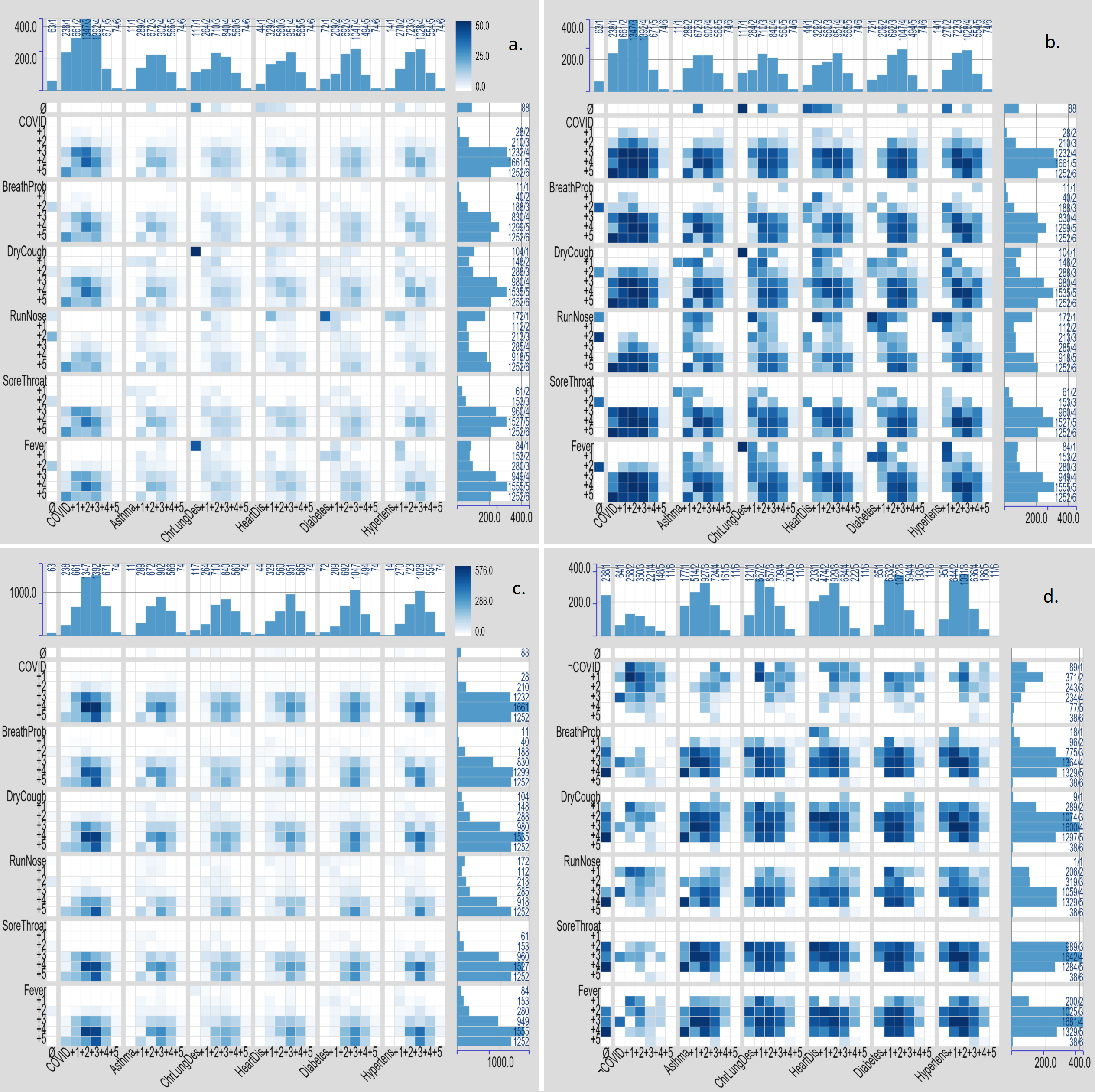}
\caption{
\textbf{a.} The main heatmap for the patient data. The \textit{COVID} variable is added to both sets, displayed on the left\,/\,top.  Item-centric value-based encoding is used, each patient is counted once, independent of the number of symptoms\,/\,diseases. 
\textbf{b.} The same data as shown in \textbf{a} but with rank-based encoding. Relative patterns are better resolved. 
\textbf{c.} Element-based encoding, each disease\,/\,symptom is counted once, a patient can contribute a value larger than 1 to the heatmap and histograms. 
\textbf{d.} The same encoding as used in \textbf{b}, but the \textit{COVID} indicator is negated: $\neg$Covid is included instead of Covid.
}
\label{fig:covid_four}
\end{figure*}
\par
We start with an analysis of co-occurrence patterns between the two sets: 
\textit{symptoms} and \textit{diseases}.
Are some combinations of symptoms and diseases particularly common? 
Is there a symptom or disease which acts different from the others? 
\par 
We start with an overview and check patterns in the heatmap as well as in the marginal histograms to drill down into interesting cases. 
Figure~\ref{fig:covid_four} shows four different modes of the top-level heatmap. 
We consider an item-centric perspective, first, counting patients (Figure~\ref{fig:covid_four}a). 
We immediately spot four isolated dark cells. They show patients with chronic lung disease (\textit{ChrLungDes}) and \textit{Fever}, \textit{ChrLungDes} and dry cough (\textit{DryCough}), \textit{ChrLungDes} without symptoms, and \textit{Diabetes} with running nose (\textit{RunNose}). 
All four cells are in the left most column of the corresponding heatmap, indicating they have \textit{ChrLundDes} only (or \textit{Diabetes} only). 
As we are showing \textit{COVID} as a set element, as well, all patients that have a single symptom or disease (the first column or row in the heatmaps) do not have \textit{COVID}. 
In the \textit{COVID}+2 column we see further dark cells. They appear in the +4 rows for all symptoms. \textit{RunNoes} has the lightest color, indicating that the combination of all other symptoms is more common.
Apparently, a running nose can be seen as an outlier, as it does not have so many items in its +1, +2, and +3 rows. Looking at the marginal histogram, we see a lot of items in the \textit{RunNoes}-alone row. Besides \textit{RunNose}, only \textit{DryCough} and \textit{Fever} show up alone; these patients have no Covid. 
\par 
Due to a large spread and a non-even distribution of the values (counts), we see many light colored cells (Fig.~\ref{fig:covid_four}a).  
Changing to a rank-based visualization (Fig.~\ref{fig:covid_four}b) reveals additional, relative patterns. 
The co-occurrence structures seem to be more coherent per disease than per symptom, for example, and for all diseases their co-occurrence with symptom \textit{RunNose} seems to be different from those to all others. 
\par 
\newcommand{\Lrot}{\raisebox{-0.25mm}{\rotatebox[origin=c]{180}{\textsf{L}}}}
Figure~\ref{fig:covid_four}c shows the data in element-centric mode~-- the sums of the histogram bins and the heatmap cells correspond to the number of symptoms and diseases.  
An \Lrot{}-shaped pattern in \textit{COVID}, \textit{Asthma}, and hypertension (\textit{Hypertens}) catches our attention for all rows, except for \textit{RunNose}. The upper heatmap rows are also less populated (patients tend to have more symptoms at once). In terms of diseases, i.e., columns, the middle area is more populated, indicating that most patients have 3, 4, or 5 diseases. 
\par 
Figure~\ref{fig:covid_four}d shows the data after inverting \textit{COVID}, focusing the analysis on patients without Covid. 
A rank-based mapping is used, as in Fig.~\ref{fig:covid_four}b. 
Asthma appears as an outlier when considering patients without Covid. The \textit{Diabetes} heatmap shows only one cell in the first row, \textit{Diabetes}+4, different from the others.  
All patients with Covid and no other disease are shown in the left-most column.  
The empty top-most row confirms that there were no Covid patients without symptoms.  
\begin{figure} 
\centering
\includegraphics[width=\linewidth]{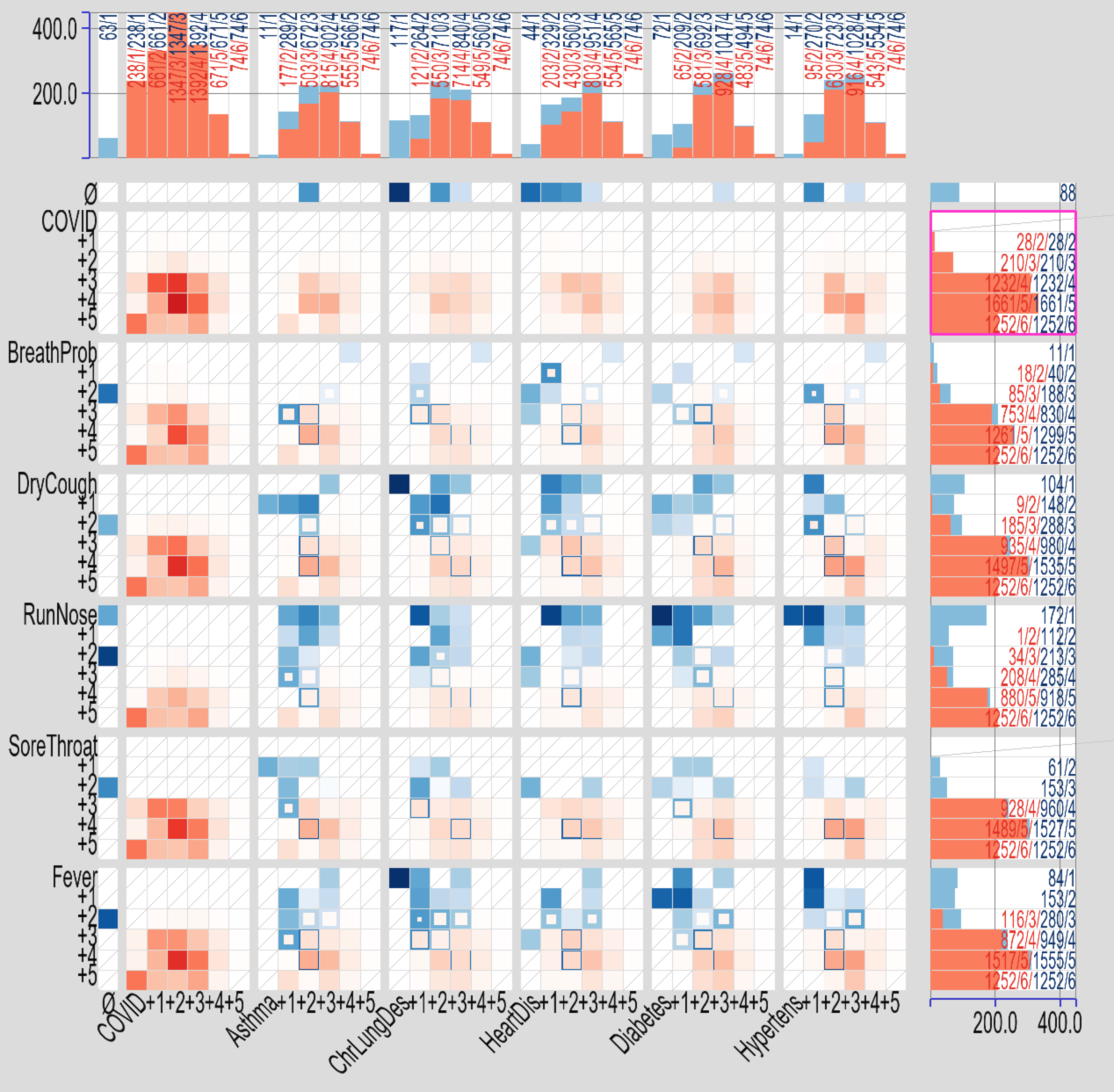}
\caption{Covid patients are selected in the marginal histogram. The top rows in the heatmaps are rarely highlighted, indicating that Covid patients have more symptoms at once.}
\label{fig:covid_brush_covid}
\end{figure}
\begin{figure}[t]
\centering
\includegraphics[width=\linewidth]{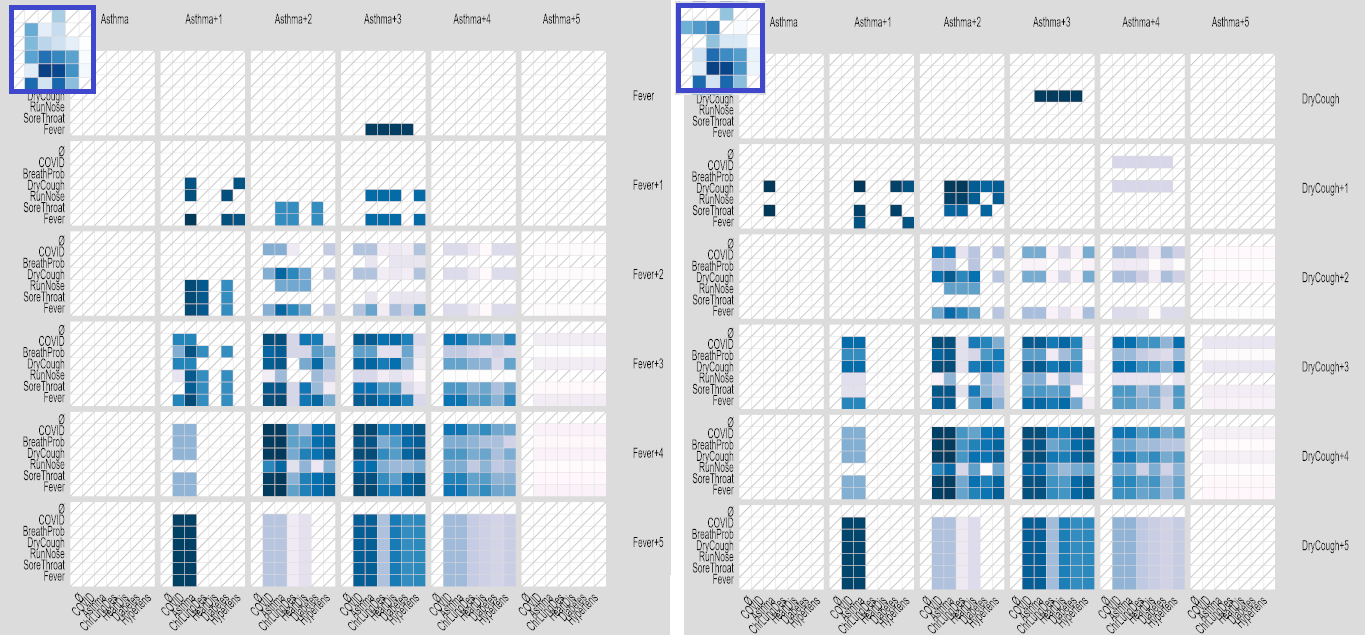}
\caption{Detail heatmaps for \textit{Asthma} vs.\ \textit{Fever} (left) and \textit{Asthma} vs. \textit{DryCough} (right) make it possible to compare co-occurrence patterns for different combinations of diseases and symptoms.}
\label{fig:covid_compare}
\end{figure}
\par 
After a first exploration of trends and outliers, we continue with a more detailed analysis. 
We select Covid patients in the marginal histogram (Fig.~\ref{fig:covid_brush_covid}). 
At large, patients with just one or two symptoms do not have Covid.
With a sore throat, at least two more symptoms are needed to co-occur with \textit{COVID}; then, however, the co-occurrnce is strong.
A few interesting exceptions pop out. In \textit{Astma}+1, we see quite some non-selected patients for the \textit{symptoms}+3 row, which is not the case for other diseases. Also \textit{DryCough}+2 is more prevalent in Covid patients than other \textit{symptoms}+2. 
\par 
Detailed histograms and heatmaps can be examined by selecting interesting parts of the heatmap.  
As we identified \textit{Asthma} as an interesting case, we examine if there is a difference in co-occurrences for patients with \textit{Asthma}\,/\,\textit{Fever} and those with \textit{Asthma}\,/\,\textit{DryCough}. 
Figure~\ref{fig:covid_compare} shows the detail heatmap for these cases.
In the case of a single symptom, both heatmaps show that it happens for \textit{Asthma}+\textit{ChrLungDes}+\textit{HeartDis}+\textit{Diabetes}. No other combination of diseases results in dry cough only, or fever only. 
In the \textit{symptom}+1 row we see some differences. \textit{DryCough} and \textit{SoreThroat} are possible for \textit{Asthma} only, and there is no pair of symptoms that includes \textit{DryCough} in case of \textit{Asthma} only. 
There are also large differences in the \textit{Asthma}+2 and \textit{Asthma}+3 columns for \textit{symptom}+1. 
The rows which show each \textit{symptom}+3 and \textit{symptom}+4 are more similar (as expected).
Still, there is a large difference in \textit{Asthma}+1 vs.\ \textit{DryCough}+3, compared to \textit{Asthma}+1 vs.\ \textit{Fever}+3. In the case of \textit{DryCough} the +1 can only be Covid, in the case of \textit{Fever}, +1 can be Covid, lung disease or Diabetes.
The +5 rows have to be same as they both include the same symptoms---all six of them.
%
%
%
\end{document}